\documentclass{article}

\usepackage[final]{neurips_2025}

\usepackage[utf8]{inputenc} 
\usepackage[T1]{fontenc}    
\usepackage{hyperref}       
\usepackage{url}            
\usepackage{booktabs}       
\usepackage{amsfonts}       
\usepackage{nicefrac}       
\usepackage{microtype}      
\usepackage[pdftex]{graphicx}
\usepackage{amsfonts}
\usepackage{amsmath}
\usepackage{enumitem}

\usepackage[table]{xcolor}
\definecolor{lightpink}{rgb}{1.0, 0.9, 0.9}

\title{Context-Aware Regularization with Markovian Integration for Attention-Based Nucleotide Analysis}

\author{%
  Mohammadsaleh Refahi \\
  Drexel University \\
  Philadelphia, PA \\
  \And
  Mahdi Abavisani \\
  Dataminr \\
  New York, NY \\
  \And
  Bahrad A. Sokhansanj \\
  Drexel University \\
  Philadelphia, PA \\
  \And
  James R. Brown \\
  Drexel University \\
  Philadelphia, PA \\
  \And
  Gail Rosen \\
  Drexel University \\
  Philadelphia, PA\\
}

\begin{document}
\maketitle
\begin{abstract}

Transformers have revolutionized nucleotide sequence analysis, yet capturing long‑range dependencies remains challenging. Recent studies show that autoregressive transformers often exhibit Markovian behavior by relying on fixed-length context windows for next-token prediction. However, standard self-attention mechanisms are computationally inefficient for long sequences due to their quadratic complexity and do not explicitly enforce global transition consistency.

We introduce \textsc{CARMANIA} (Context-Aware Regularization with Markovian Integration for Attention-Based Nucleotide Analysis), a self-supervised pretraining framework that augments next-token (NT) prediction with a transition-matrix (TM) loss. The TM loss aligns predicted token transitions with empirically derived n-gram statistics from each input sequence, encouraging the model to capture higher-order dependencies beyond local context. This integration enables \textsc{CARMANIA} to learn organism-specific sequence structures that reflect both evolutionary constraints and functional organization.

We evaluate \textsc{CARMANIA} across diverse genomic tasks, including regulatory element prediction, functional gene classification, taxonomic inference, antimicrobial resistance detection, and biosynthetic gene cluster classification. \textsc{CARMANIA} outperforms the previous best long-context model by at least 7\%, matches state-of-the-art on shorter sequences (exceeding prior results on 20/40 tasks while running $\sim$2.5$\times$ faster), and shows particularly strong improvements on enhancer and housekeeping gene classification tasks—including up to a 34\% absolute gain in Matthews correlation coefficient (MCC) for enhancer prediction. The TM loss boosts accuracy in 33 of 40 tasks, especially where local motifs or regulatory patterns drive prediction. This enables more effective modeling of sequence-dependent biological features while maintaining robustness across non-coding and low-signal regions. Code available at \url{https://github.com/EESI/carmania}.

\end{abstract}

\section{Introduction}

Deoxyribonucleic acid (DNA), often referred to as the "language of life," encodes the genetic instructions essential for the development, functioning, and reproduction of all known living organisms and many viruses. The sequence of its four nucleotide bases—adenine (A), thymine (T), cytosine (C), and guanine (G)—forms the foundation of genetic information, dictating the synthesis of proteins and the regulation of various biological processes.
Advancements in high-throughput sequencing technologies have exponentially increased genomic data availability.

Despite advances in sequencing, analyzing large-scale biological data remains challenging. Traditional methods—such as motif detection, sequence alignment, and Markov models—capture short-range dependencies well~\cite{delcher2007identifying,steinegger2017mmseqs2,wood2019improved}, but often fail to represent complex or long-range patterns, especially in diverse or non-canonical sequences.
Recent work has applied NLP techniques to biological sequences~\cite{nguyen2024hyenadna,cahyawijaya2022snp2vec,schiff2024caduceus,bo2025revisiting}, with large language models (LLMs) showing strong ability to capture complex dependencies. However, genomic sequences are often extremely long, making full attention computationally infeasible. To address this, transformer models use mechanisms like sliding window and sparse attention~\cite{dong2025hymba,zaheer2020big}, which expand the receptive field but may still struggle to capture long-range dependencies effectively in genomics.


In this work, we enhance transformers by incorporating Markovian priors, guiding them to follow long-range transition dynamics and better capture complex dependencies within the broader context of genomic sequences.
Drawing on recent studies that interpret transformers' capabilities through the lens of Markovian processes~\citet{hu2024limitation,zekri2024large}, we propose a novel pretraining strategy that explicitly enforces these Markovian properties.

Our approach, named \textsc{CARMANIA} (Context-Aware Regularization with Markovian Integration for Attention-Based Nucleotide Analysis), supplements the standard next-token prediction objective with an auxiliary loss. This loss aligns the model’s predicted first-order transition matrix with one derived from the full input sequence, encouraging consistency with its empirical token transition patterns. By doing so, \textsc{CARMANIA} captures the probabilistic structure of genomic sequences and improves context-awareness during next-token prediction. 


To scale pretraining to long genomic sequences, we replace full attention with \textbf{sliding-window attention (SWA)}, a sparse mechanism that limits each token’s receptive field to a fixed-size window of preceding tokens, reducing complexity from $O(n^2)$ to $O(kn)$ ($k \ll n$) while preserving biologically meaningful local interactions. \textbf{TM loss} complements this by reinforcing short-range dependencies within the local window while promoting global consistency, since the transition matrix is computed over the entire input sequence. Together, these components enable a balance between local motif recognition and broader statistical alignment. \emph{In practice, CARMANIA handles effective contexts up to 160\,kbp, which makes it, to our knowledge, the longest-context \textbf{transformer-based genomic language model} to date.} An overview of our  framework is illustrated in Figure~\ref{fig:experiments}.

In addition to improving context modeling, our pretraining approach supports efficient domain adaptation, as conserved transition patterns across species enable generalization with minimal fine-tuning~\cite{bergeron2023evolution}.

\noindent The main contributions of this work are:

\begin{figure}[t]
\centering
  \includegraphics[width=1\columnwidth]{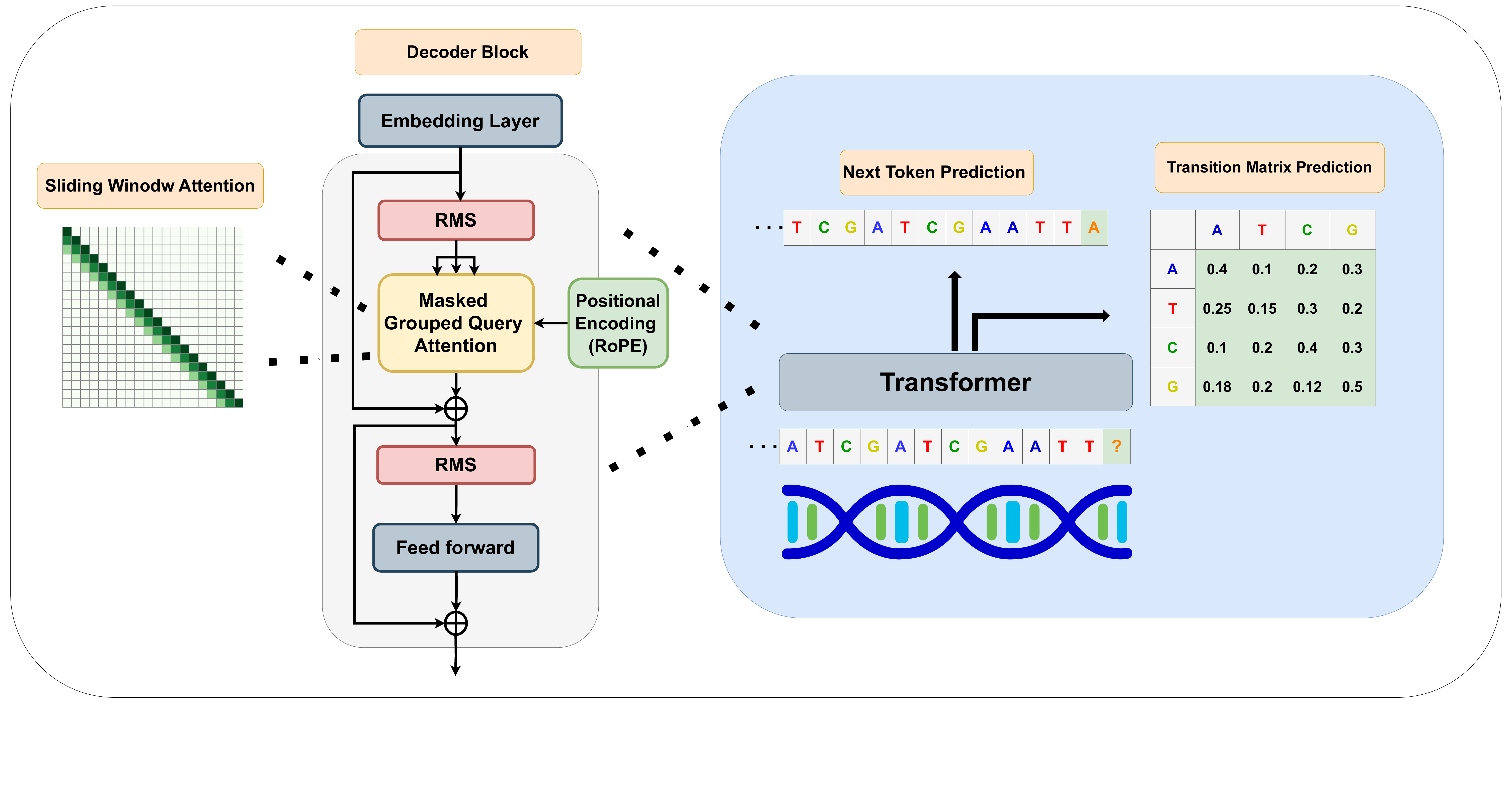}
  \vspace{-1.1cm}
\caption{\textbf{Proposed Pretraining Framework.} We extend a LLaMA-style decoder with a transition matrix module to capture global nucleotide co-occurrence. The model uses sliding-window attention with rotary embeddings and local caching, reducing attention complexity from $O(n^2)$ to $O(n)$. The transition matrix complements local attention by preserving long-range dependencies efficiently.}
  \label{fig:experiments}
  \vspace{-0.1cm}
\end{figure}

\textbf{• A scalable architecture for long-context genomic modeling} that integrates sliding-window attention, RoPE-based positional encoding, FlashAttention, and a widened Transformer backbone, enabling efficient representation learning over extended DNA sequences. \\

\textbf{• A novel self-supervised objective based on Transition Matrix (TM) loss}, which regularizes the transformer model by aligning predicted token transitions with empirical n-gram distributions. This Markovian prior enhances memory retention, out-of-distribution generalization, and biological coherence across long genomic contexts. \\

\textbf{• Comprehensive evaluation across 40 genomic tasks}, including regulatory element prediction, functional gene classification, taxonomic inference, and biosynthetic gene cluster(BGC) prediction, where our model achieves consistent improvements—even in benchmarks where alternative inductive biases have previously dominated.

\section{Related Work}

\subsection{Markovian Models in Genomics}

Hidden Markov Models (HMMs) are widely used in genomics for modeling the probabilistic structure of biological sequences. They enable gene prediction, as in GeneMark~\cite{lukashin1998genemark}, and other methods for identifying coding regions~\cite{delcher2007identifying}. Profile HMMs are commonly applied to protein family analysis, capturing conserved regions and domain variability~\cite{potter2018hmmer}. HMMs also power microbial classification tools like Clasnip~\cite{chuan2023clasnip}, and Aphmm~\cite{firtina2024aphmm} improves the speed and efficiency of profile HMM computations. Their effectiveness depends on model structure, sequence length, and the biological complexity of the task.

\subsection{Language Models in Genomics}
The first breakthrough in genomic language models came with DNABERT~\cite{ji2021dnabert}, which employed a BERT architecture with k-mer tokenization and sequence masking to learn DNA sequence representations. The Nucleotide Transformer~\cite{dalla2024nucleotide} scaled this concept, training a 2.5 billion parameter model on 850 genomes. Other models like HyenaDNA~\cite{nguyen2024hyenadna} used long convolutional blocks for efficient long-sequence handling, while BigBird~\cite{refahi2024scorpio,refahi2023leveraging} applied sparse and sliding window attention to capture long-range dependencies. Caduceus~\cite{schiff2024caduceus}, built on Mamba, introduced a bi-directional model preserving reverse complement symmetry.

Together, these advances in tokenization, architecture, and representation learning have led to increasingly effective models for genomic analysis~\cite{cahyawijaya2022snp2vec,celikkanat2024revisiting}.

\subsection{The Markovian Core of Autoregressive Transformers}

There has been significant research \citet{edelman2024evolution,zekri2024large,ildiz2024selfattentionmarkovmodelsunveiling,hu2024limitation} aimed at establishing an equivalence between autoregressive language models based on transformers and Markov chains. Specifically, findings emphasize the role of statistical induction heads, which adaptively learn in-context patterns, forming a foundation for next-token predictions. This perspective introduces a compelling connection between autoregressive architectures and Markov chains, paving the way for a nuanced understanding of in-context learning (ICL).\citet{zekri2024large,edelman2024evolution}.
\citet{zekri2024large} demonstrates that an autoregressive model with a vocabulary size \(T\) and a context window of size \(K\) can be represented as a Markov chain with a state space of size \(O(T^K)\). 

In an autoregressive model, the probability of predicting the next token \(x_{t+1}\) given the previous \(K\) tokens \((x_t, x_{t-1}, \ldots, x_{t-K+1})\) can be written as:
\begin{equation}
\Pr(x_{t+1} \mid x_t, x_{t-1}, \ldots, x_{t-K+1}).
\label{eq:markov_autoreg}
\end{equation}

This conditional probability can be interpreted as a transition probability in a $K$-th order Markov chain, where each unique $K$-length token sequence defines a state. The transition matrix \(Q \in \mathbb{R}^{T \times T}\) is defined as:
\begin{equation}
Q_{s,s'} = \Pr(x_{t+1} = x_{s'} \mid x_t = x_s),
\label{eq:transition_matrix}
\end{equation}
where \(s\) and \(s'\) index sequences of \(K\) tokens. The rows of \(Q\) sum to 1, satisfying the Markov property. Thus, the model's behavior approximates a Markov chain with \(T^K\) states, capturing dependencies within the fixed context window.
Although self-attention is globally contextual and non-Markovian in theory, limited context in practice causes autoregressive Transformers to approximate Markovian behavior~\cite{ildiz2024selfattentionmarkovmodelsunveiling}.

\section{Pretraining Long Sequences With Markovian Knowledge}

We adopt a causal Transformer inspired by LLaMA\cite{touvron2023llama}, where each layer computes hidden representations $\mathbf{h}^{(l)}$ from input tokens $\mathbf{x} = (x_1, x_2, \dots, x_n)$ and the previous layer’s outputs $\mathbf{h}^{(l-1)}$, using multi-head self-attention to predict the next token. To scale to long sequences, we replace full self-attention with SWA, where each token attends only to a fixed-size $k$ window of preceding tokens, ${x_{t-k}, \dots, x_{t-1}}$, reducing complexity from $O(n^2)$ to $O(kn)$. This not only improves efficiency but also reflects the local nature of most biological dependencies in genomic sequences. We further improve efficiency with FlashAttention-2~\cite{dao2023flashattention} and enhance positional encoding via Rotary Positional Embeddings (RoPE)~\cite{su2024roformer}, enabling better extrapolation to long contexts.

\subsection{Unified $n$-th Order Transition Tensor Formulation}

To complement the next-token prediction objective, we introduce a unified probabilistic framework that captures higher-order dependencies through a transition tensor. Let $\mathbf{P}_t \in \mathbb{R}^V$ denote the model's predicted probability distribution over the vocabulary $\mathcal{V}$ at position $t$, where $[\mathbf{P}_t]_i = \mathbb{P}(x_t = i \mid x_{<t})$. 

We define the $n$-th order transition tensor $\mathcal{T}^{(n)} \in \mathbb{R}^{V^n}$ as:
\begin{equation}
    \mathcal{T}^{(n)}_{i_1 i_2 \dots i_n} = \frac{1}{B(L-n+1)} \sum_{b=1}^{B} \sum_{t=1}^{L-n+1} \prod_{s=0}^{n-1} \left[ \mathbf{P}_{t+s}^{(b)} \right]_{i_{s+1}},
\end{equation}
where $B$ is the batch size, $L$ is the sequence length, $V$ is the vocabulary size, and $i_1, \dots, i_n$ index token values in the $n$-gram.

This formulation approximates the joint distribution over $n$ consecutive model predictions, providing a global summary of sequential dynamics. In practice, we focus on the first-order case ($n = 2$), which corresponds to modeling bigram transitions. This strikes a balance between expressiveness and computational efficiency, and directly supports the TM loss introduced in the next section.

\subsection{Pretraining Objective}

Our pretraining objective combines two complementary components: next-token prediction and transition matrix alignment. The primary objective is the standard autoregressive language modeling loss, which minimizes the negative log-likelihood of predicting each token \( x_t \) given its preceding context:
\begin{equation}
    \mathcal{L}_{\text{NT}} = -\sum_{t=1}^n \log P_\theta(x_t \mid x_1, \dots, x_{t-1}),
\end{equation}
where \( P_\theta \) denotes the model’s predicted token distribution.

To reinforce global sequence structure, we introduce a Transition Matrix (TM) loss that encourages the model to match the empirical token transition dynamics. Let \( p_{ij} \) and \( q_{ij} \) denote the smoothed n-gram distributions of the input and model outputs, respectively. The KL divergence between these distributions defines the TM loss:
\begin{equation}
    \mathcal{L}_{\text{TM}} = \sum_{i,j} p_{ij} \log \frac{p_{ij}}{q_{ij}}.
\end{equation}

The full loss combines both terms:
\begin{equation}
    \mathcal{L}_{\text{Full}} = \mathcal{L}_{\text{NT}} + \beta \mathcal{L}_{\text{TM}},
\end{equation}
where \( \beta \) is a tunable hyperparameter controlling the influence of the TM loss.

\section{Experiments}
In this section, we evaluate \textsc{CARMANIA} across a range of genomic tasks. We first describe the pretraining datasets, model architecture, and training setup, followed by ablation studies assessing the impact of the TM loss and attention mechanism on convergence, efficiency, and long-range retention. Finally, we present comprehensive benchmark results demonstrating the model’s adaptability, generalization, and superior performance over existing methods.

\begin{figure}[t]
  \centering
  \includegraphics[width=0.44\linewidth]{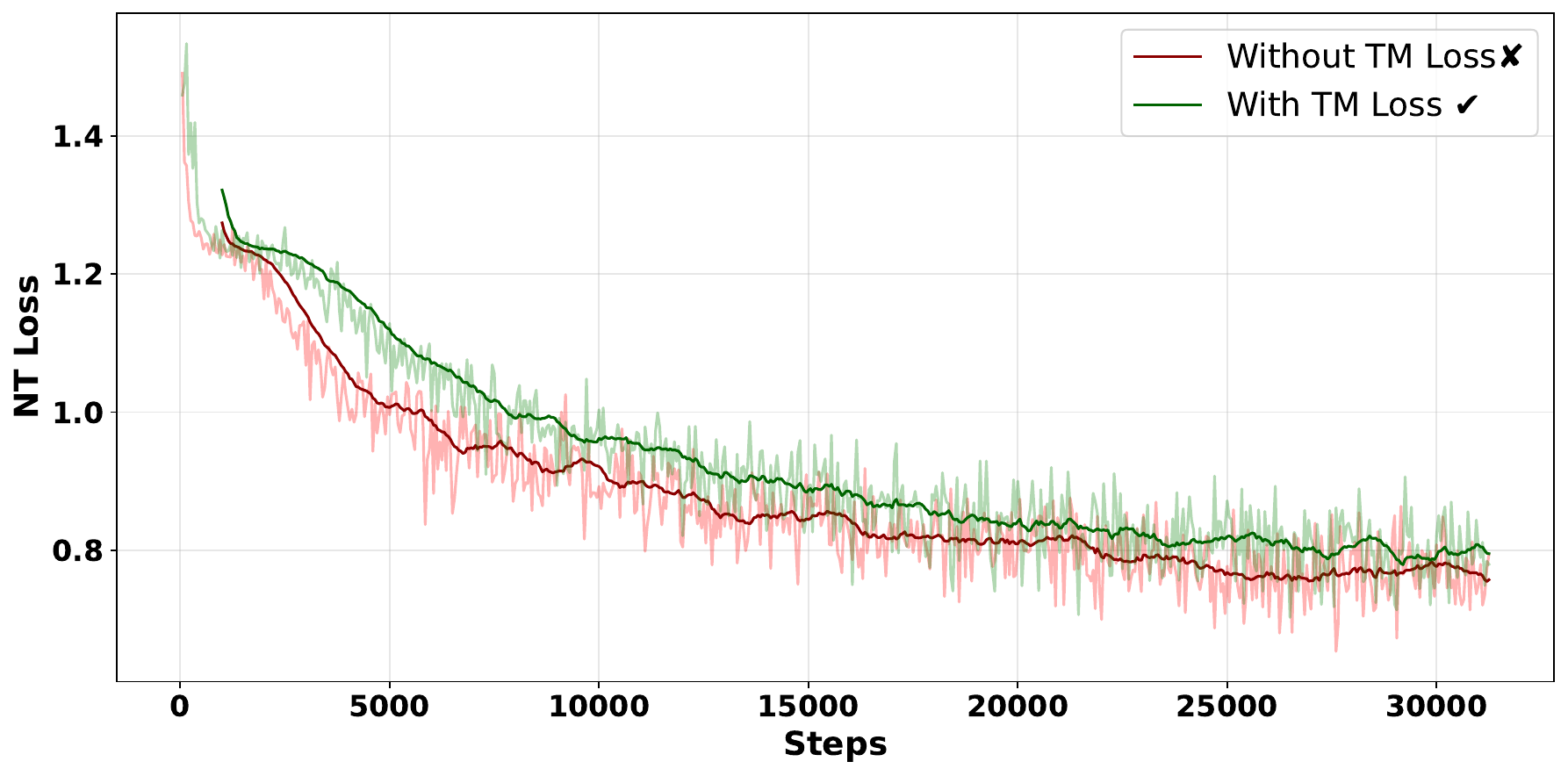} \hfill
  \includegraphics[width=0.44\linewidth]{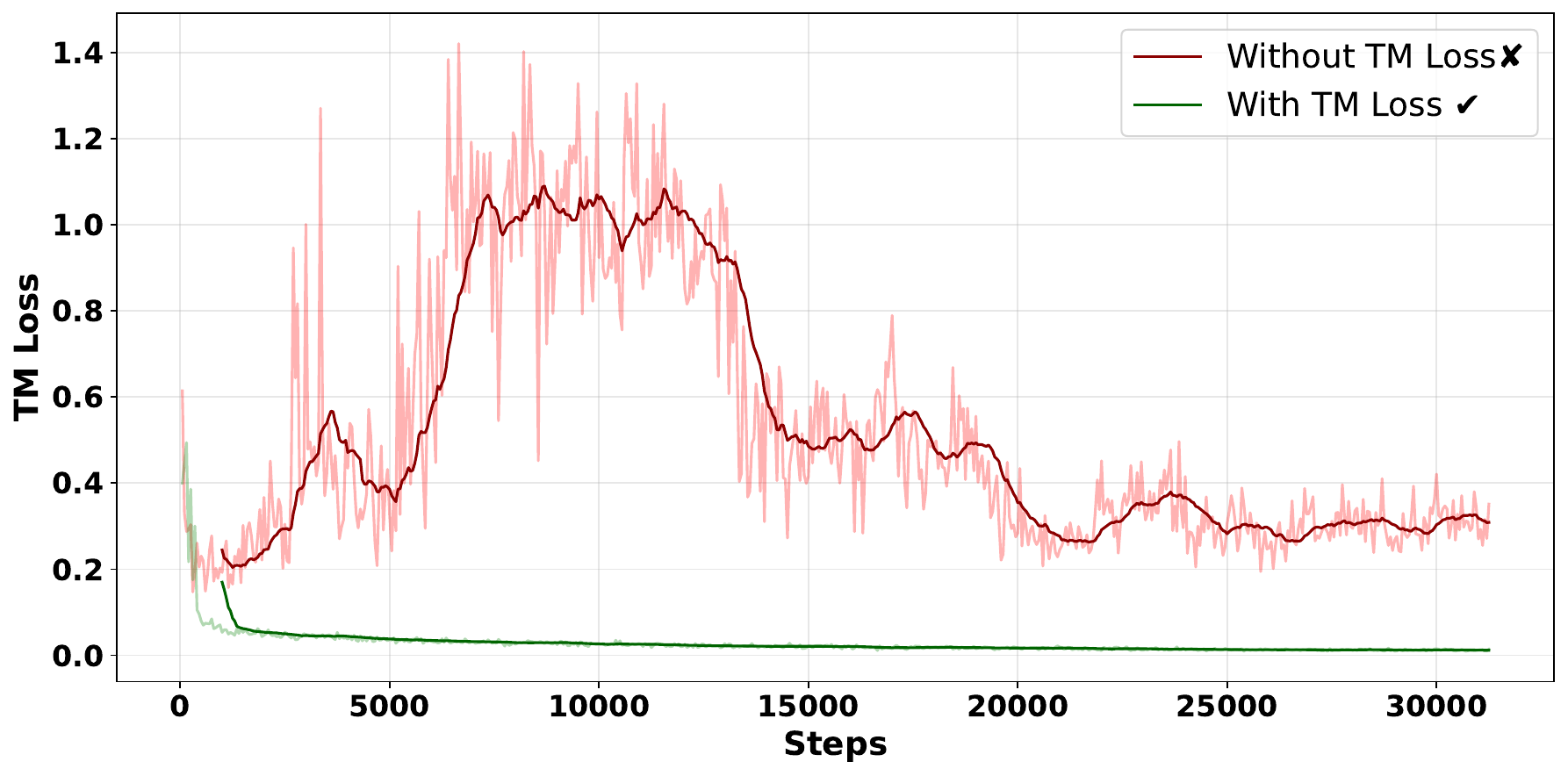}
  \caption{Comparison of training losses with and without explicit TM loss enforcement ($\beta=1$ vs. $\beta=0$).  
\textbf{Left:} Both models show similar reductions in next-token prediction loss.  
\textbf{Right:} Without TM loss ($\beta=0$), the model partially learns transition structure—TM loss rises then stabilizes—indicating implicit alignment with n-gram patterns.}
  \label{fig:comp_tm}
\end{figure}


\begin{table}[t]
\centering
\caption{
Ablation study showing the impact of TM loss and attention type on model performance ({F1 Macro}, {BLEU}) and compute cost. 
All models are trained on the {Scorpio-Gene–Taxa} dataset. 
{F1 Macro} is evaluated on the {Antimicrobial Resistance (AMR)} classification tasks, while {BLEU} and {Relative FLOPs} are computed on the {Human Genome–Short} dataset using 10{,}000 fragments from the {GRCh38/hg38} reference genome. 
Relative FLOPs are reported with respect to the {Sliding Window} attention baseline, which has an absolute compute cost of approximately $16.7\times10^{12}$ FLOPs per sequence.
}

\label{tab:ablation_study}
\small
\renewcommand{\arraystretch}{1.0}
\resizebox{0.85\linewidth}{!}{
\begin{tabular}{lccc}
\toprule
\textbf{Model Setting} & \textbf{F1 Macro (↑)} & \textbf{BLEU (↑)} & \textbf{Relative FLOPs (↓)} \\
\midrule
Transformer (SWA)                          & 0.873 ± 0.130 & 0.73 & 1.00 \\
+ TM Loss                                  & 0.882 ± 0.131 & 0.77 & 1.00 \\
+ Full Attention + TM Loss                 & 0.883 ± 0.131 & 0.71 & 1.58 \\
\bottomrule
\end{tabular}
}
\vspace{-5pt}
\end{table}

\subsection{Experimental Setup} 

\subsubsection*{Pre-training Datasets}

We pre-train our model on three large-scale genomic datasets: (i) GRCh38~\cite{grch382013p13}, which provides $\sim$3B base pairs with 10~kbp and 160~kbp fragments; (ii) the Basic Genome Dataset~\cite{zhu2022phylogeny}, consisting of $\sim$10B base pairs across 4,600+ genomes with 10~kbp fragments; and (iii) the Scorpio Gene-Taxa Dataset~\cite{refahi2024scorpio}, comprising $\sim$580M base pairs from 2,046 genomes with 4~kbp fragments. Additional details are in Appendix~A.

\subsubsection*{Model and Training Details}
\noindent\textbf{Architecture}:
\textsc{CARMANIA} is a LLaMA-based causal Transformer tailored for genomic sequence modeling. It uses 16 attention heads (4 key-value), a window size of 128, and 5 custom Transformer layers with an embedding size of 1024 and intermediate dimension of 4608. The model uses SiLU activations and contains 83M parameters. We selected this wide architecture based on the comparative results in Supplementary Table~\ref{tab:deep_vs_wide_model}, which demonstrate its superior performance and efficiency over deeper models in both in-domain and out-of-domain genomic tasks.

\noindent\textbf{Tokenization and n-gram Frequency Calculation:}
We tokenize DNA sequences at the single-nucleotide level (A, T, C, G) to preserve fine-grained features such as SNPs and avoid disrupting open reading frames, as can occur with subword or non-overlapping k-mer methods. Additionally, during tokenization, we compute a normalized $4 \times 4$ first-order transition matrix for each sequence based on bigram frequencies, where each row defines a probability distribution over possible next nucleotides. This sequence-specific matrix acts as a self-supervised training signal, serving as ground truth for the transition-matrix loss and guiding the model to learn biologically meaningful transition patterns.

\noindent\textbf{Training Parameters, and Hardware Specifics}:
The model was trained for two epochs using PyTorch on an NVIDIA A100 GPU (80GB). To fit within memory constraints, batch sizes were set based on sequence length: 35 for 4~kbp, 19 for 10~kbp, and 1 for 160~kbp inputs. We used a cosine annealing schedule with an initial learning rate of 5e-4. Additional training details, including optimizer settings and gradient clipping, are listed in Supplementary Table~\ref{tab:param_ranges}. The training setup yields an efficiency of approximately $7.57 \times 10^{-9}$ GPU-hours per token.

\begin{figure}[t]
  \centering
  \includegraphics[width=0.52\linewidth]{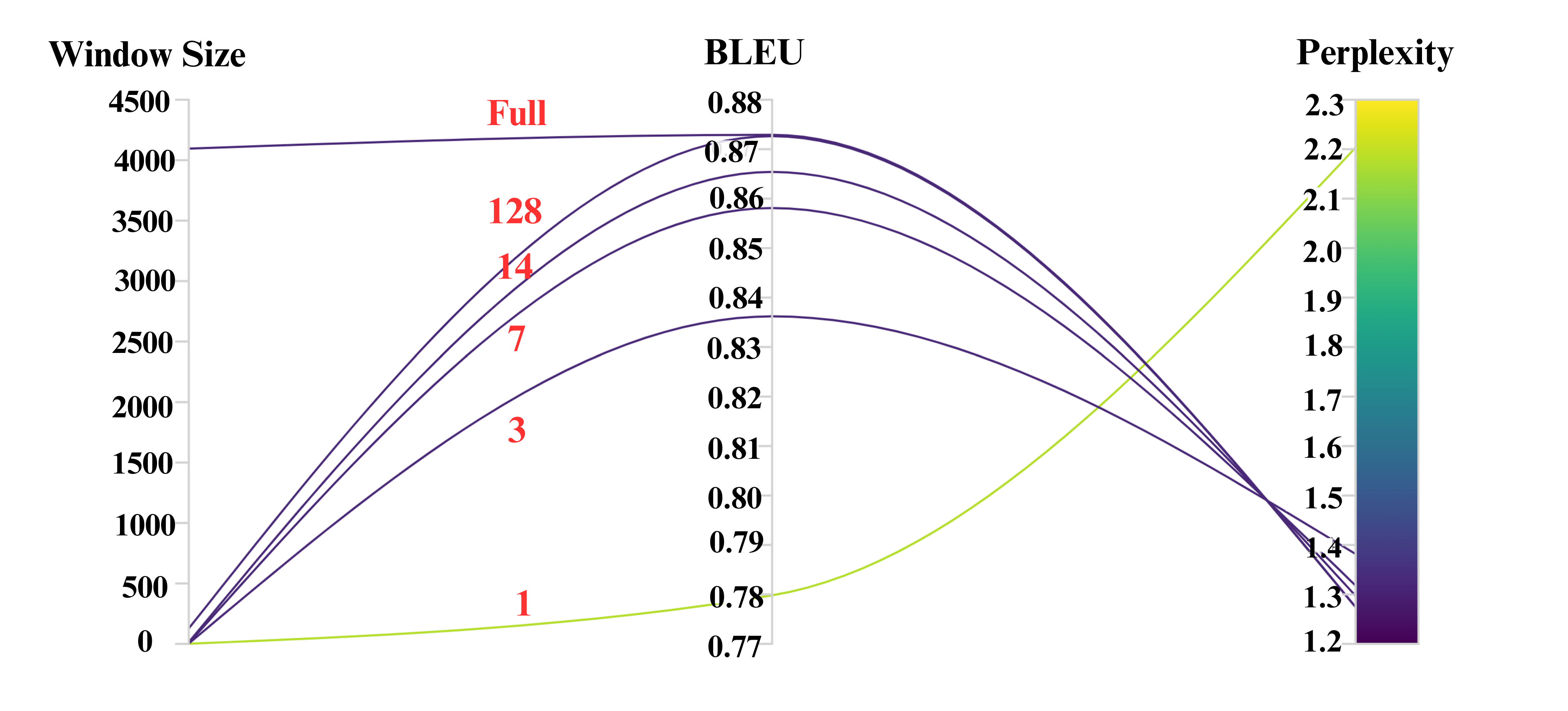} \hfill
  \includegraphics[width=0.42\linewidth]{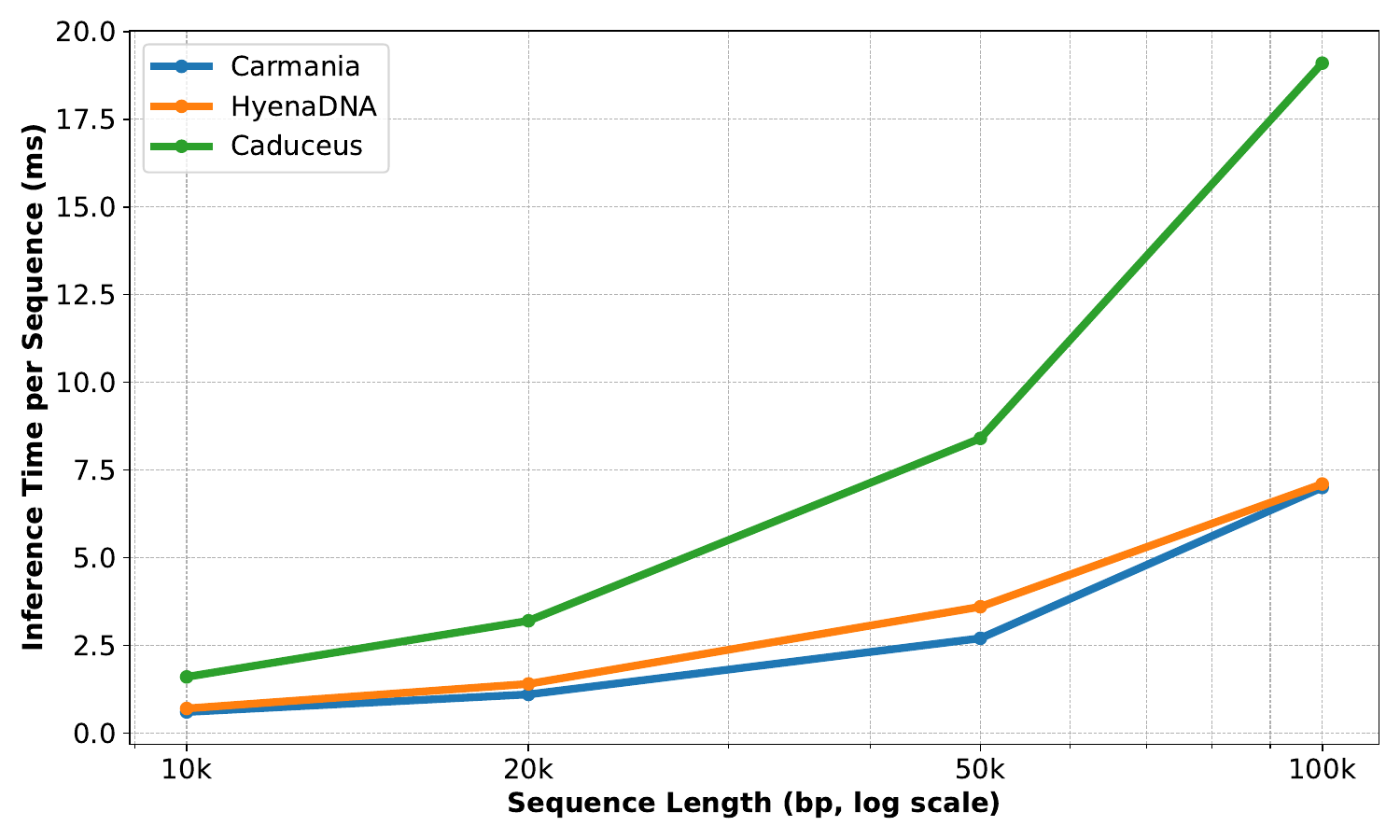}
  \caption{\textbf{Left:} Effect of window size on model performance. A window size of 128 achieves results comparable to full attention.
    \textbf{Right:}Inference time per sequence for \textsc{CARMANIA}(83M), HyenaDNA(1.6M), and Caduceus-PH(1.9M) across varying sequence lengths.
  }
\label{fig:window_effect}
\vspace{-5pt}
\end{figure}

\subsection{Ablation Studies}

\subsubsection*{Effect of Transition Matrix Loss}
We study the effect of adding first-order TM loss alongside the next-token (NT) loss by training on the Scorpio-Gene-Taxa dataset for 31k steps under two settings: with TM loss ($\beta=1$) and without it ($\beta=0$). In the $\beta=0$ case, TM loss is computed but not used during training, allowing us to observe the model’s natural transition behavior.The left panel of Figure~\ref{fig:comp_tm} shows NT loss decreases similarly for both models, indicating that adding TM loss does not hinder token-level learning. In contrast, the right panel reveals different TM loss patterns: for the $\beta=0$ model, the TM loss initially increases then decreases as the model partially and passively captures transition structure, while for the $\beta=1$ model, the TM loss consistently decreases from the start. However, as shown in supplementary results, convolution-based models like HyenaDNA require explicit TM regularization to exhibit this property.

Finally, as reported in Table~\ref{tab:ablation_study}, models trained with TM loss achieve higher macro F1 scores on downstream Antimicrobial Resistance (AMR) classification tasks (detailed in Supplementary Table~\ref{tab:finetune_stats}) and improved BLEU scores on the human genome benchmark. We also evaluated this effect across all 40 downstream benchmarks and found that adding TM loss led to improved performance in 33 of them.

\noindent\textbf{Higher-Order Transition Matrices.}  
As shown in Supplementary Tables~\ref{tab:f1_macro_compare_tm_orders} and~\ref{tab:scorpio_compare_tm_orders}, applying second-order TM loss led to performance degradation across tasks compared to first-order TM or no TM. This drop is likely due to the sparsity of higher-order transitions in biological sequences, which makes the loss signal unstable. In contrast, first-order TM loss consistently improves performance by capturing reliable and biologically meaningful pairwise dependencies.

\subsubsection*{Effect of Windowed vs. Full Attention}
We assess the effect of attention window size by training on a 100M bp subset of the Scorpio-Gene-Taxa dataset. As shown in Figure~\ref{fig:window_effect}, larger windows improve BLEU scores and reduce perplexity, with full attention as an upper bound. Notably, a window size of 128 performs comparably to full attention, indicating that TM loss enables smaller windows to capture complex dependencies efficiently. Figure~\ref{fig:window_effect} also illustrates inference time comparisons across models on input sequences of varying lengths, up to 100k bases. Despite having 83M parameters, our model runs approximately 2.5 times faster than Caduceus becomes significantly slower as sequence length increases, while HyenaDNA (1.6M) maintains relatively fast inference but underperforms on long sequences and remains slower than our model. This demonstrates the practical efficiency of our architecture, and also highlights how windowed attention enables faster inference in Transformer-based models.

Table~\ref{tab:ablation_study} compares the performance of models trained with full attention and windowed attention. While full attention achieves slightly better performance on downstream tasks, it results in a lower BLEU score.
In addition, we evaluate the computational efficiency of windowed attention. As shown in Table~\ref{tab:ablation_study}, full attention requires 58\% more computational resources compared to a window size of 128. This demonstrates the trade-off between accuracy and efficiency, as windowed attention reduces complexity from \(O(n^2)\) in full attention to \(O(k^2n)\), making it a more practical and scalable choice for large-scale sequence modeling.

\subsection{Long-Range Sequence Retention}
We evaluate each model’s ability to preserve long-range sequence structure by measuring internal consistency across extended human genome regions. Specifically, we analyze 50 independent 160 kbp genomic segments. Within each segment, we extract 100 bp windows using a fixed stride of 2000 bp and compare the model-predicted sequences against the corresponding regions in the original input sequence using Hamming similarity. This produces a position-wise similarity profile, which we average across all segments to assess how well each model maintains fidelity over increasing genomic distance.

\begin{figure}[t]
  \centering
  \includegraphics[width=0.45\linewidth]{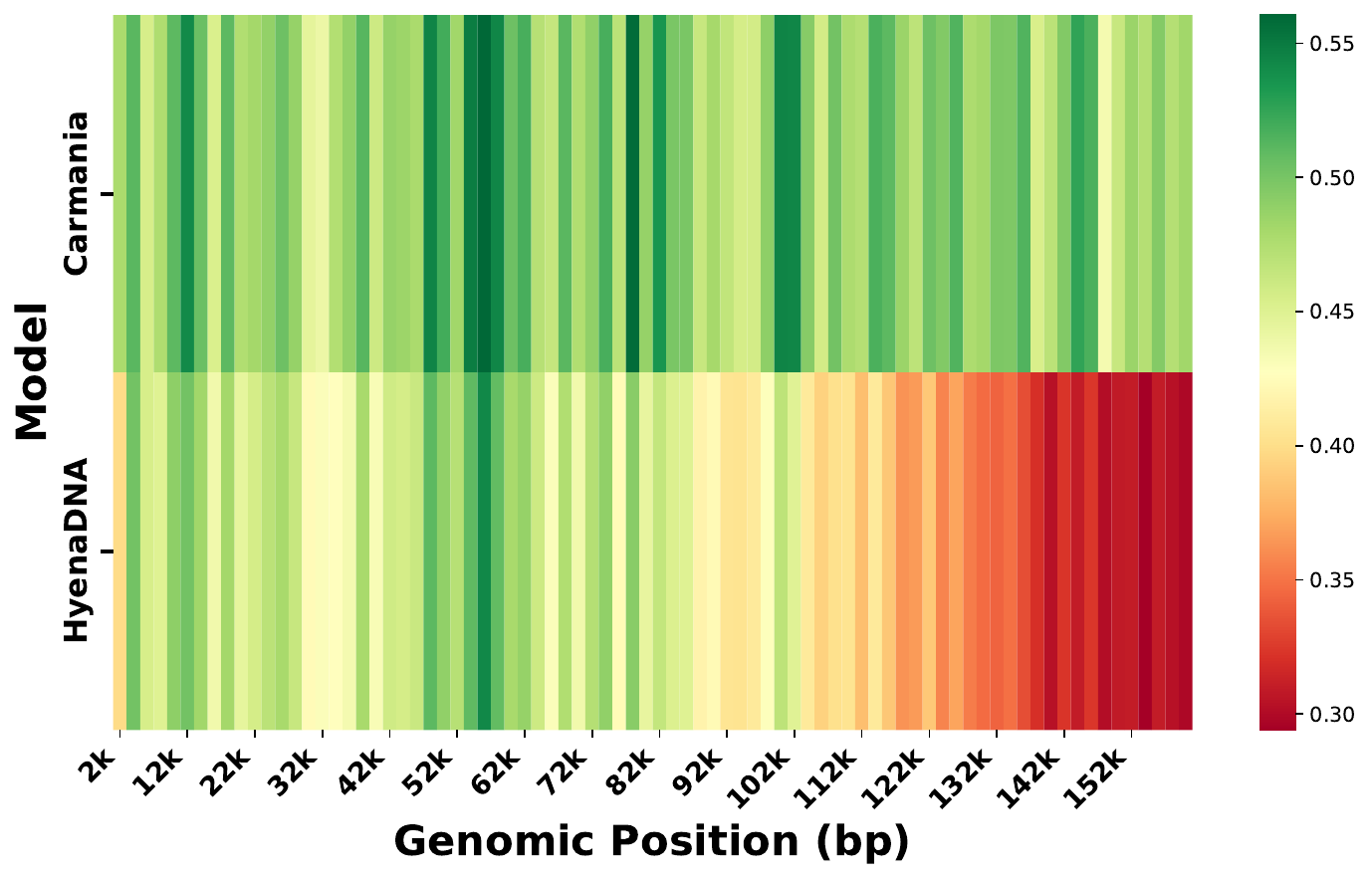} \hfill
  \includegraphics[width=0.45\linewidth]{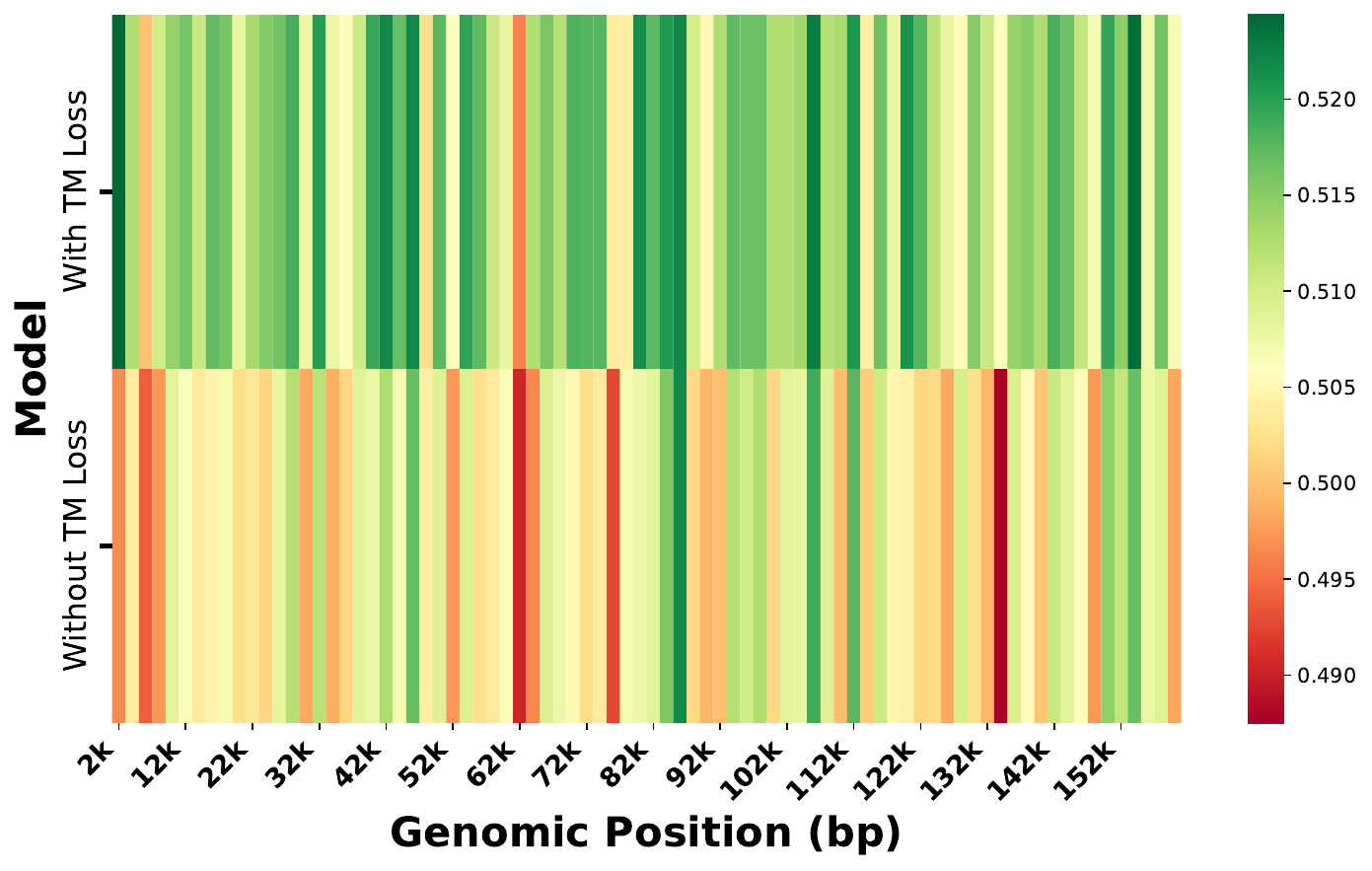}
  \caption{
    Heatmap of average sequence similarity across 50 independent 160~kbp genomic segments in 100~bp windows at 2000~bp intervals.\textbf{Left:} \textsc{CARMANIA} consistently maintains high similarity across all regions, demonstrating superior long-range memory, whereas HyenaDNA shows reduced sequence coherence in later segments.
    \textbf{Right:} Incorporating the TM loss improves memory retention in \textsc{CARMANIA}, yielding higher sequence similarity across extended contexts.
  }
\label{fig:heatmap_combined}
\end{figure}

As shown in Figure~\ref{fig:heatmap_combined}, HyenaDNA shows a marked drop in sequence similarity at distal positions. This likely reflects the limitations of its fixed-depth recurrence architecture, where information must propagate layer by layer without explicit token-to-token interaction \cite{nguyen2024hyenadna}, making it harder to retain detailed signals over long spans. In contrast, \textsc{CARMANIA} maintains high similarity across the entire sequence due to its sliding-window attention and the TM loss, which encourages consistency in token transitions.
The effect of the TM loss is further illustrated in Figure~\ref{fig:heatmap_combined}, where models trained without it show noticeably shorter retention spans, suggesting its role in supporting full-sequence memorization.

\subsection{Downstream Task Evaluations}

\subsubsection*{Genomics Benchmark: Supervised Adaptation}

To evaluate task-specific adaptation, we fine-tuned our model on diverse genomics classification datasets. The Genomic Benchmarks collection includes tasks such as regulatory element prediction, enhancer detection, and binary species classification~\cite{grevsova2023genomic}.
For a fair comparison, we fine-tuned the human-pretrained \textsc{CARMANIA} model using 5-fold cross-validation and compared its performance to other state-of-the-art models trained on the human genome. As shown in Table~\ref{tab:genome_bencmakr_comparison}, \textsc{CARMANIA} achieves top performance in 4 out of 8 tasks and matches the best performance in 2 others, demonstrating its strong generalization ability in supervised downstream settings.

\begin{table}[h]
\centering
\caption{
Top-1 accuracy (↑) across 5-fold cross-validation on Genomic Benchmark tasks. Results for CNN, HyenaDNA, Mamba, and both Caduceus variants are taken from the original Caduceus paper~\citep{schiff2024caduceus}.  \underline{Underlined} values indicate second-best results.
}
\label{tab:genome_bencmakr_comparison}
\resizebox{\textwidth}{!}{
\begin{tabular}{lccccc|cc}
\toprule
\textbf{Task} & \textbf{CNN } & \textbf{HyenaDNA} & \textbf{Mamba}  & \textbf{Caduceus-PH} & \textbf{Caduceus-PS } & \textbf{\textsc{CARMANIA}}& \textbf{Baseline w/o TM}\\
\midrule
Mouse Enhancers & 0.715 ± 0.087 & \underline{0.780 ± 0.025} & 0.743 ± 0.054 &  0.754 ± 0.074 & \textbf{0.793 ± 0.058} & 0.761 ± 0.019 & 0.748 ± 0.020\\
Coding vs. Intergenic & 0.892 ± 0.008 & 0.904 ± 0.005 & 0.904 ± 0.004 &  0.915 ± 0.003 & 0.910 ± 0.003  & \textbf{0.935 ± 0.001} & \underline{0.930 ± 0.001}\\
Human vs. Worm & 0.942 ± 0.002 & 0.964 ± 0.002 & 0.967 ± 0.002 &  \textbf{0.973 ± 0.001} & \underline{0.968 ± 0.002} & \underline{0.968 ± 0.003} & 0.966 ± 0.001\\
Human Enhancers Cohn & 0.702 ± 0.021 & 0.729 ± 0.014 & 0.732 ± 0.029 &  \textbf{0.747 ± 0.004} & \underline{0.745 ± 0.007} & 0.724  ± 0.005 & 0.699  ± 0.005\\
Human Enhancer Ensembl & 0.744 ± 0.122 & 0.849 ± 0.006 & 0.862 ± 0.008 &  0.893 ± 0.008 &\underline{ 0.900 ± 0.006} & \textbf{0.916 ± 0.002}  & 0.892 ± 0.001  \\
Human Regulatory & 0.872 ± 0.005 & 0.869 ± 0.012 & 0.814 ± 0.211 &  0.872 ± 0.011 & 0.873 ± 0.007 & \textbf{0.895 ± 0.002} & \underline{0.893 ± 0.002}\\
Human OCR Ensembl & 0.698 ± 0.013 & 0.783 ± 0.007 & 0.815 ± 0.002 &  \textbf{0.828 ± 0.006} & \underline{0.818 ± 0.006} & 0.775 ± 0.002 & 0.763 ± 0.002\\
Human NonTATA Promoters & 0.861 ± 0.009 & 0.944 ± 0.002 & 0.933 ± 0.007 &  0.946 ± 0.007 & 0.945 ± 0.010 & \textbf{0.963 ± 0.002} & \underline{0.961 ± 0.003} \\
\bottomrule
\end{tabular}}
\end{table}

\subsubsection*{Nucleotide  Transformer Tasks: Supervised Adaptation}

We benchmarked our model on 18 diverse genomics classification tasks introduced by Dalla-Torre et al.~\cite{dalla2024nucleotide}, spanning histone modification prediction, regulatory annotation, and splice site detection. Following their evaluation protocol, we performed 10-fold cross-validation with early stopping, reporting the mean and standard deviation across different random seeds.


As shown in Table~\ref{tab:histone_regulatory_annotation}, \textsc{CARMANIA} achieves top-1 performance on 5 out of 18 tasks and outperforms the average of all baselines by over 3\% across metrics. Notably, in enhancer classification tasks, \textsc{CARMANIA} achieves up to 34\% absolute improvement in MCC over previous models \cite{zhou2023dnabert,avsec2021effective,schiff2024caduceus,nguyen2024hyenadna,dalla2024nucleotide}. These results demonstrate the model’s strong ability to capture regulatory signals from sequence.
\noindent
A notable gain is observed on the H4ac histone mark, where adding the TM loss yields over 40.9\% relative improvement in MCC. The benefit of the TM loss, however, is not uniform across all genomics tasks. In enhancer prediction, where positive and negative classes differ strongly in their local motif distributions (KL divergence = 9.28), the TM loss provides no measurable improvement ($\Delta = 0.0\%$). In splice-site donor prediction (KL = 0.99), it yields a modest 5.2\% gain, and in the H4ac histone mark dataset (KL = 0.64), a substantial 40.9\% improvement. These results demonstrate that TM-driven performance gains inversely track inter-class bigram separability, confirming that TM loss is most effective when local sequence features are ambiguous and long-range dependencies become critical. Biologically, this aligns with our understanding that enhancer regions possess clear, unique motifs that NT loss can easily learn, whereas splice sites rely on short, dispersed sequence cues. In contrast, H4 acetylation involves multiple lysine sites (K5, K8, K12, K16) on the H4 N-terminal tail that act cooperatively to regulate chromatin accessibility~\cite{dion2005genomic,ma2024chhm}. Instead of strong individual motifs, these sites follow diffuse and combinatorial sequence patterns, making TM’s global co-occurrence modeling particularly advantageous.

\begin{table}[h]
\centering
\caption{
Performance comparison (↑) on 10-fold cross-validation across 18 tasks, including histone marks, regulatory elements, and splice site prediction. We report MCC for histone/enhancer tasks, F1 for promoter/splice site, and accuracy for the “All” splice task. Baselines follow Caduceus~\cite{schiff2024caduceus} and Dalla-Torre et al.~\cite{dalla2024nucleotide}. \textsc{CARMANIA} achieves top performance on 5 tasks and remains competitive elsewhere. \underline{Underlined} values denote second-best results.}
\label{tab:histone_regulatory_annotation}
\resizebox{\textwidth}{!}{
\begin{tabular}{lcccccc|cc}
\toprule
\textbf{} & \textbf{\textbf{ENFORMER}} & \textbf{\textbf{DNABERT-2}} & \textbf{\textbf{NT-V2}} & \textbf{\textbf{HYENADNA}} & \textbf{\textbf{Caduceus-PH}} & \textbf{\textbf{Caduceus-PS}} & \textbf{\textbf{\textsc{CARMANIA}}} & \textbf{\textbf{Baseline w/o TM}} \\
 & 252M & 117M & 500M & 1.6M & 1.9M & 1.9M & 83M &  83M \\
\midrule
\textbf{Histone Markers} \\
H3 & 0.719±0.048 & 0.785±0.033 & 0.784±0.047 & 0.779±0.037 &  \textbf{0.815±0.048} &  \underline{0.799±0.029} & 0.782 ±0.023 & 0.785 ±0.011  \\
H3K14AC & 0.288±0.077 & 0.516±0.028 & 0.551±0.021 & 0.612±0.065 & \textbf{0.631±0.026} & 0.541±0.212 & \underline{0.627 ±0.021}  & 0.631 ±0.022  \\
H3K36ME3 & 0.344±0.055 & 0.591±0.020 & 0.625±0.013 & 0.613±0.041 & 0.601±0.129 & 0.609±0.109 & \textbf{0.632 ±0.011}& \underline{0.629 ±0.017}  \\
H3K4ME1 & 0.291±0.061 & 0.511±0.028 & \textbf{0.550±0.021} & 0.512±0.024 & \underline{0.523±0.039} & 0.488±0.102 & 0.515 ±0.017 &0.516 ±0.017  \\
H3K4ME2 & 0.211±0.069 & 0.336±0.040 & 0.319±0.045 & 0.455±0.095 & 0.487±0.170 & 0.388±0.101 & \textbf{0.502 ±0.025}  & \underline{0.500 ±0.038} \\
H3K4ME3 & 0.158±0.072 & 0.352±0.077 & 0.410±0.033 & 0.549±0.056 & 0.544±0.045 & 0.440±0.202 & \underline{0.565 ±0.012} & \textbf{0.586 ±0.011}  \\
H3K79ME3 & 0.496±0.042 & 0.613±0.030 & 0.626±0.026 & 0.672±0.048 & \underline{0.697±0.077} & 0.676±0.026 & \textbf{0.699 ±0.013} & 0.695 ±0.018 \\
H3K9AC & 0.420±0.063 & 0.542±0.029 & 0.562±0.040 & 0.581±0.061 & \textbf{0.622±0.030} & 0.604±0.048 & \underline{0.615 ±0.013} & 0.608 ±0.010  \\
H4 & 0.732±0.076 & 0.796±0.027 & \underline{0.799±0.025} & 0.763±0.044 & \textbf{0.811±0.022} & 0.789±0.020  & 0.772±0.010   & 0.769±0.019  \\
H4AC & 0.273±0.063 & 0.463±0.041 & 0.495±0.032 & 0.564±0.038 & \textbf{0.621±0.054} & 0.525±0.240 & \cellcolor{lightpink}\underline{0.606 ±0.014}  & \cellcolor{lightpink}0.197±0.011 \\
\midrule
\textbf{Regulatory Annotation} \\
ENHANCER & 0.451±0.108 & 0.516±0.098 & 0.548±0.144 & 0.517±0.117 & 0.546±0.073 & 0.491±0.066 & \cellcolor{lightpink} \textbf{0.880±0.013} &  \cellcolor{lightpink}\underline{0.878±0.013}  \\
ENHANCER TYPES & 0.309±0.134 & 0.423±0.051 & 0.424±0.132 & 0.386±0.185 & 0.439±0.054 & 0.416±0.095 & \cellcolor{lightpink}\textbf{0.724±0.013} &\cellcolor{lightpink} \underline{0.724±0.010} \\
PROMOTER: ALL & 0.954±0.006 & 0.971±0.006 & \textbf{0.976±0.006} & 0.960±0.005 & \underline{0.970±0.004} & 0.967±0.004 & 0.963±0.001 & 0.960±0.001 \\
\space  NONTATA & 0.955±0.010 & 0.972±0.005 & \textbf{0.976±0.005} & 0.959±0.008 &  \underline{0.969±0.011} & 0.968±0.006 & 0.962±0.002 & 0.960±0.003\\
\space   TATA & 0.960±0.023 & 0.955±0.021 &  \textbf{0.966±0.013} & 0.944±0.040 & 0.953±0.016 &  \underline{0.957±0.015 } & 0.942±0.008 & 0.940±0.008  \\
\midrule
\textbf{Splice Site Annotation} \\
ALL & 0.848±0.019 & 0.939±0.009 & \textbf{0.983±0.008} & 0.956±0.011 & 0.940±0.027 & 0.927±0.021 &  \underline{0.967±0.008} & 0.967±0.004  \\
ACCEPTOR & 0.914±0.028 & 0.975±0.006 & \textbf{0.981±0.011} & 0.958±0.010 & 0.937±0.033 & 0.936±0.077 & \underline{0.958±0.005} & 0.943±0.010 \\
DONOR & 0.906±0.027 & 0.963±0.006 & \textbf{0.985±0.022} & 0.949±0.024 & 0.948±0.025 & 0.874±0.289 & \underline{0.973±0.002} & 0.921±0.018 \\
\bottomrule
\end{tabular}
}
\vspace{-5pt}
\end{table}

\subsubsection*{Task-Adaptive Pre-Training on the Scorpio-Gene-Taxa Dataset}

We evaluated task-adaptive pre-training on the \textit{Scorpio-Gene-Taxa} dataset~\citep{refahi2024scorpio}, which includes three evaluation splits: \textit{Test} (seen genes and taxa), \textit{Gene-out} (unseen gene labels), and \textit{Taxa-out} (unseen phyla). All models were trained on the same data and evaluated using FAISS-based~\citep{douze2024faiss} embedding similarity to retrieve the best match from the training set.

As shown in Table~\ref{tab:scorpio_compare}, \textsc{CARMANIA} achieves the highest accuracy on both the main test set and the \textit{Taxa-out} split, while also showing strong generalization on the \textit{Gene-out} set. In contrast, MetaBERTa performs well on \textit{Gene-out} but struggles on \textit{Taxa-out}, suggesting it may rely more on taxonomic memorization than on learning a transferable mapping from gene sequence to taxonomy. This highlights a fundamental challenge in this dataset: there is often a trade-off between generalizing across gene identities and taxonomic groups. Most models struggle to do both simultaneously unless they explicitly capture the hierarchical structure linking genes to taxa~\citep{refahi2024scorpio}.

\begin{figure*}[t]
    \centering
    \raisebox{1.2ex}{\small \textbf{\textsc{CARMANIA}}} \hspace{0.1\linewidth}
    \raisebox{1.2ex}{\small \textbf{HyenaDNA}} \hspace{0.1\linewidth}
    \raisebox{1.2ex}{\small \textbf{MetaBERTa(BigBird)}} \\
    
    \includegraphics[width=0.26\linewidth]{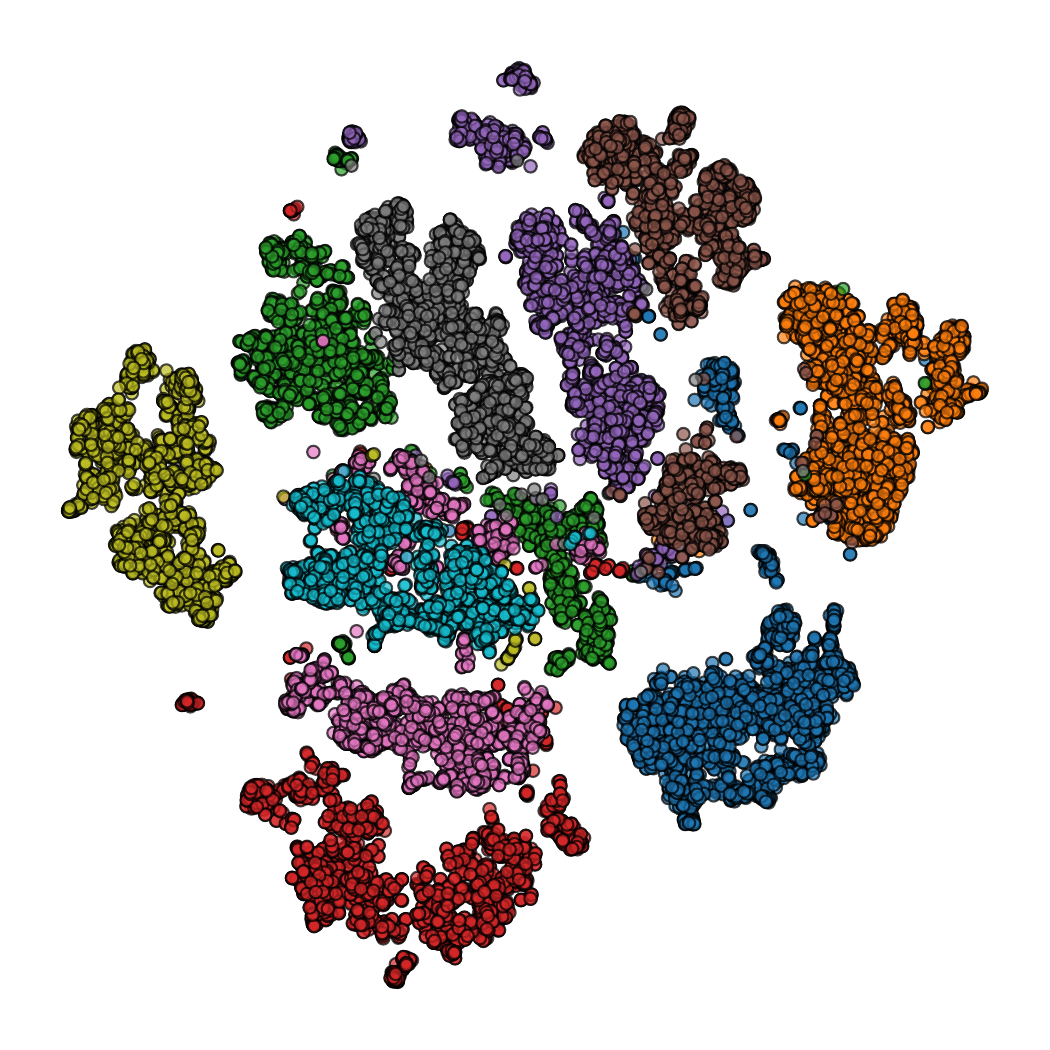} 
    \includegraphics[width=0.26\linewidth]{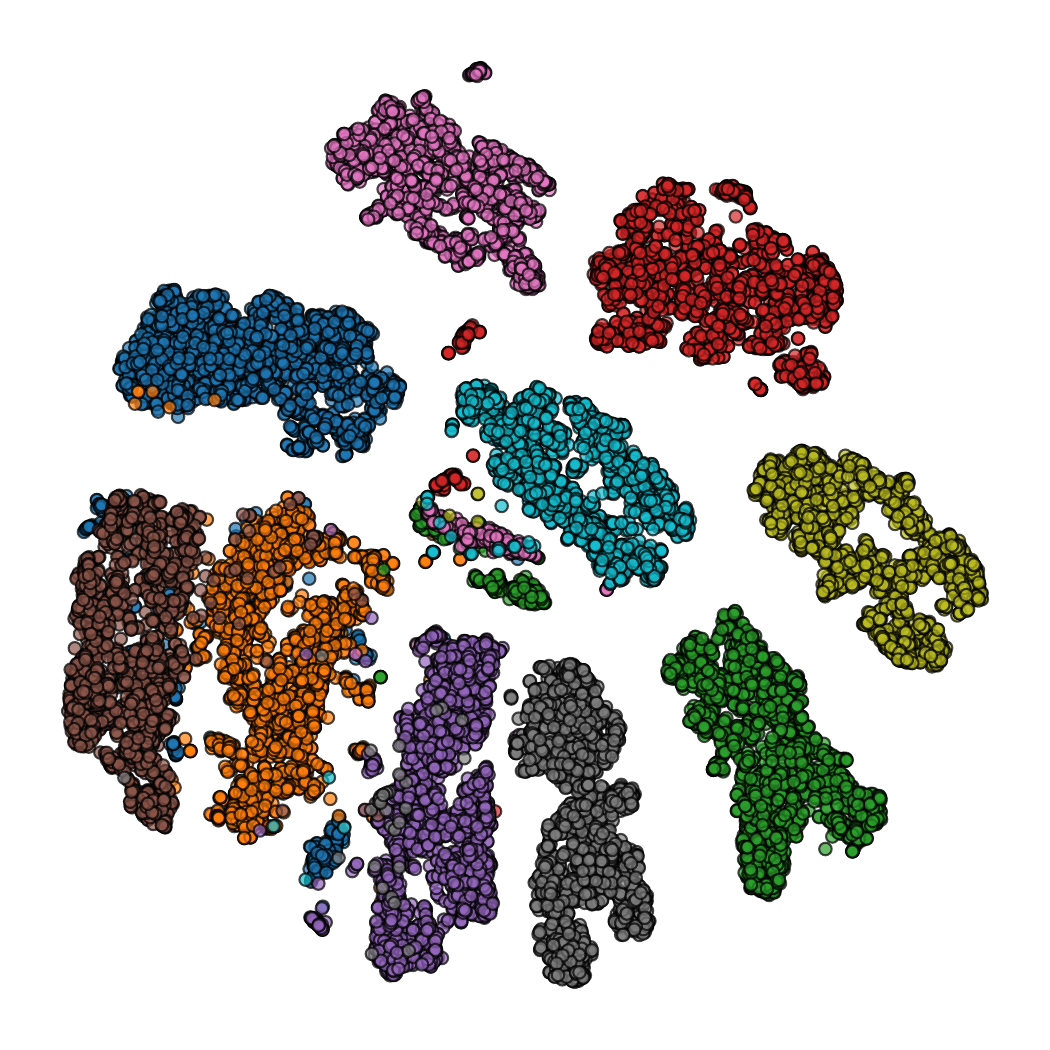}
    \includegraphics[width=0.26\linewidth]{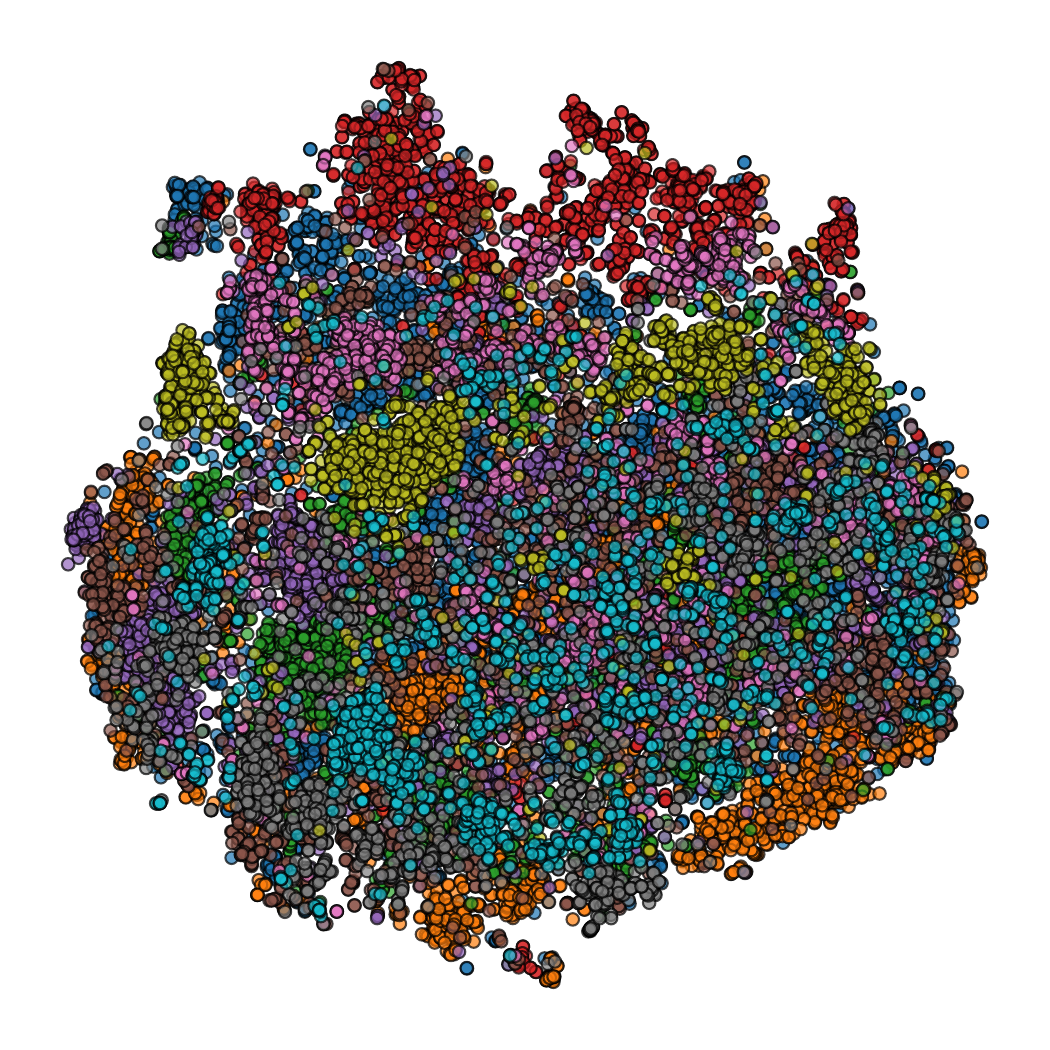}
    \centerline{\includegraphics[width=0.85\linewidth]{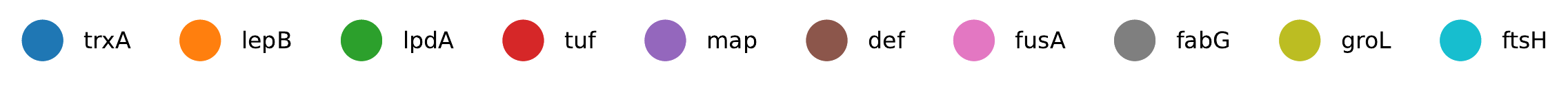}}
\caption{t-SNE visualization of the 10 most common genes in the Scorpio-Gene-Taxa dataset. \textsc{CARMANIA} effectively clusters genes while maintaining taxonomic coherence, leading to superior gene-to-taxonomy classification performance.}
\label{fig:tsne_comparison}
\end{figure*}

To further examine these results, we visualized the learned embeddings using t-SNE for the 10 most frequent genes (Figure~\ref{fig:tsne_comparison}). \textsc{CARMANIA} forms compact clusters that align well with both gene identity and taxonomic structure, suggesting it captures both levels of organization. HyenaDNA produces well-separated clusters based on gene identity but lacks alignment with taxonomic labels. In contrast, MetaBERTa embeddings show stronger alignment with taxonomy but weaker separation by gene, as also seen in Supplementary Figure~\ref{fig:tsne_comparison_phylum}.

This is further supported by additional results in Supplementary Table~\ref{tab:f1_macro_compare}, where \textsc{CARMANIA} outperforms all other models on AMR gene classification tasks without fine-tuning, highlighting its strong domain adaptation capability.

\begin{table}[t]
\centering
\caption{Performance comparison on the Scorpio-Gene-Taxa dataset.}
\label{tab:scorpio_compare}
\resizebox{0.87\textwidth}{!}{
\begin{tabular}{lccccc|cccc|cc}
\toprule
& \multicolumn{5}{c}{\textbf{Test}} & \multicolumn{4}{c}{\textbf{Gene out}} & \multicolumn{1}{c}{\textbf{Taxa out}} \\
\hline
& \textbf{Phylum} & \textbf{Class} & \textbf{Order} & \textbf{Family} & \textbf{Gene} & \textbf{Phylum} & \textbf{Class} & \textbf{Order} & \textbf{Family} & \textbf{Gene} \\
\hline
MetaBERTa & 0.712 & 0.584 & 0.426 & 0.322 & 0.282  & \textbf{0.640} & \textbf{0.471} & \textbf{0.290} & \textbf{0.204} & 0.074 \\
HyenaDNA & 0.764 & 0.636 & 0.447 & 0.292 & 0.846 & 0.449 & 0.265 & 0.113 & 0.058 & 0.674 \\
Baseline w/o TM & 0.860 & 0.765 & 0.589 & 0.417 & 0.845 & 0.547 & 0.401 & 0.222 & 0.145 & 0.608  \\
\textsc{CARMANIA} & \textbf{0.861} & \textbf{0.768} & \textbf{0.596} & \textbf{0.419} & \textbf{0.909} & 0.469 & 0.316 & 0.159 & 0.094 & \textbf{0.728} \\
\bottomrule
\end{tabular}}
\end{table}


\subsubsection*{Domain Adaptation and Generalization}

We evaluated several pre-trained genomic models on the AMR dataset \underline{without} fine-tuning. The dataset, derived from MEGARes~\cite{bonin2023megares} and CARD~\cite{jia2016card}, includes three classification tasks: Gene Family, Resistance Mechanism, and Drug Class~\cite{yoo2024predicting}. Labels were assigned based on the closest match (best hit) in embedding space using FAISS\cite{douze2024faiss} on frozen representations.

As shown in Table~\ref{tab:f1_macro_compare}, \textsc{CARMANIA} achieves the highest F1 Macro scores across all tasks, indicating that it produces more informative embeddings for AMR classification than prior models, including HyenaDNA\citet{nguyen2024hyenadna}, Caduceus\citet{schiff2024caduceus}, MetaBERTa \citet{refahi2024scorpio}, and Nucleotide Transformers\citet{dalla2024nucleotide}.

\begin{table}[h]
\centering
\small
\caption{F1 Macro scores for AMR classification on Gene Family, Resistance Mechanism, and Drug Class using FAISS-based best-hit embedding retrieval.}
\label{tab:f1_macro_compare}
\renewcommand{\arraystretch}{0.95}
\resizebox{0.75\textwidth}{!}{
\begin{tabular}{lccc}
\toprule
\textbf{Model} & \textbf{GeneFamily} & \textbf{Resist-Mech} & \textbf{DrugClass} \\
\midrule
MetaBERTa (35.2M) & 0.650 & 0.898 & 0.886 \\
HyenaDNA (1.6M) & 0.623 & 0.840 & 0.880 \\
Caduceus-PH (1.9M) & 0.597 & 0.823 & 0.839 \\
NucleotideTrans (2.5B) & 0.611 & 0.856 & 0.859 \\
\midrule
w/o TM (83M) & 0.728 & 0.974 & 0.931 \\
\textsc{CARMANIA} (83M) & \textbf{0.733} & \textbf{0.975} & \textbf{0.942} \\
\bottomrule
\end{tabular}}
\vspace{-1pt}
\end{table}

\subsection*{Long Context Task: Biosynthetic Gene Cluster Classification}

To evaluate the ability of genomic language models to capture functional patterns across extended DNA regions, we performed a large-scale classification task on biosynthetic gene clusters (BGCs) using the MiBiG database~\cite{kautsar2020mibig, liu2022deep}, where each cluster is labeled by its associated secondary metabolite class. This task was framed as multi-class classification using raw DNA sequences—without relying on protein translation or domain-level annotations. Since BGCs vary in length (average ~377k bp), we truncated all sequences to 100k bp for model compatibility.

Unlike previous approaches that depend on external tools like Prodigal and HMMER~\cite{hannigan2019deep, liu2022deep}, our method directly operates on nucleotide sequences. We evaluated performance using 5-fold cross-validation. As shown in Table~\ref{tab:bgc_results}, \textsc{CARMANIA}, which incorporates TM loss, achieves the highest accuracy (48.4\%), outperforming HyenaDNA by over 7\% and significantly surpassing other baselines.

\begin{table}[h]
\centering
\small
\caption{5-fold CV accuracy and standard deviation for BGC classification on 100{,}000 DNA sequences.}
\label{tab:bgc_results}
\renewcommand{\arraystretch}{0.95}
\resizebox{0.75\textwidth}{!}{
\begin{tabular}{lcccc}
\toprule
\textbf{Model} & \textbf{Caduceus-Ph} & \textbf{HyenaDNA} &  \textbf{Baseline w/o TM} & \textbf{\textsc{CARMANIA}} \\
\midrule
Accuracy ± Std & 0.326 ± 0.012 & 0.412 ± 0.013 & 0.410 ± 0.024 & \textbf{0.484 ± 0.033} \\
\bottomrule
\end{tabular}}
\end{table}

\section{Conclusions}

In this paper, we introduced \textsc{CARMANIA}, a novel pre-trained genomic model designed for long-sequence analysis and nucleotide-level processing. Unlike existing models, \textsc{CARMANIA} effectively captures both local and global sequence dependencies, making it highly suitable for diverse genomic tasks. Our evaluations demonstrate superior performance across multiple benchmarks, including AMR classification, gene-taxa association, regulatory element prediction, species classification, and biosynthetic gene cluster prediction.
\paragraph{Limitations and Future Work:} While our method improves sequence representation learning, one limitation lies in the fixed scaling parameter $\beta$ used to balance the next-token and TM losses. We set $\beta=1$ and observed that the losses naturally aligned in scale, making this choice effective in practice. However, tasks involving rare but biologically important motifs may require more fine-grained or dataset-specific tuning. In cases where $H_{\text{emp}}(X) \neq H_{\text{model}}(X)$, a fixed $\beta$ may lead to overfitting on common patterns while missing informative outliers. Future work could explore adaptive scaling strategies that dynamically adjust $\beta$ based on the training loss distribution.
Although the TM loss is computationally efficient due to its matrix-based formulation, memory usage increases with larger vocabulary sizes and batch dimensions. This can limit scalability, particularly in domains such as chemistry (e.g., SMILES representations) or natural language, where token sets are larger and sequence variability is higher. To address this, future work could explore memory-efficient approximations, sparse matrix operations, or gradient checkpointing to reduce the computational footprint.
In addition, we aim to scale up both the model and the diversity of training datasets to improve generalization across a broader range of genomic and functional prediction tasks. Finally, we note that, like other predictive genomic models, this approach may raise concerns around potential misuse in synthetic biology or privacy-sensitive contexts, which should be addressed in future deployments.

\newpage
\bibliographystyle{plainnat}
\bibliography{neurips_2025}
\newpage

\appendix
\label{sec:appendix}

\section{Supplementary Material}


\subsection{Effect of TM Loss on Other Domains: Protein Modeling}

While CARMANIA was primarily developed for genomic sequences, we further examined the generality of the proposed \textbf{TM loss} in other biological domains. In theory, TM loss should benefit any modeling task that relies on capturing long-range dependencies. To validate this hypothesis, we adapted the CARMANIA architecture for amino acid sequences and trained it on the \textbf{Scorpio-Gene–Taxa} protein dataset. Specifically, we constructed a $20 \times 20$ bigram frequency matrix representing pairwise transition statistics between standard amino acids and incorporated TM loss into the training objective.

As summarized in Table~\ref{tab:tm_protein}, incorporating TM loss consistently accelerated convergence and improved predictive accuracy across all hierarchical levels. These results confirm that TM loss generalizes beyond nucleotide modeling and effectively enhances protein-sequence representations.

\begin{table}[h!]
\centering
\caption{Performance comparison on the \textbf{Scorpio-Gene–Taxa} dataset using protein sequences.}
\label{tab:tm_protein}
\small
\renewcommand{\arraystretch}{1.05}
\resizebox{0.9\linewidth}{!}{%
\begin{tabular}{lcccccccccc}
\toprule
 & \multicolumn{5}{c}{\textbf{Test Set}} & \multicolumn{4}{c}{\textbf{Gene-Heldout}}  & \multicolumn{1}{c}{\textbf{Taxa-Heldout}}\\
\cmidrule(lr){2-6} \cmidrule(lr){7-10}  \cmidrule(lr){11-11} 
\textbf{Model} & \textbf{Phylum} & \textbf{Class} & \textbf{Order} & \textbf{Family} & \textbf{Gene} & \textbf{Phylum} & \textbf{Class} & \textbf{Order} & \textbf{Family} & \textbf{Gene} \\
\midrule
Without TM Loss & 91.1 & 84.9 & 70.6 & 51.3 & 99.5 & 19.8 & 10.9 & 3.3 & 1.1 & 98.4 \\
With TM Loss    & 91.4 & 85.5 & 71.0 & 51.5 & 99.6 & 20.5 & 11.6 & 3.8 & 1.5 & 98.7 \\
\bottomrule
\end{tabular}%
}
\end{table}

\subsection{Effect of TM Loss on Convolution-Based Architecture: HyenaDNA}
\begin{figure}[h]
  \includegraphics[width=0.48\linewidth]{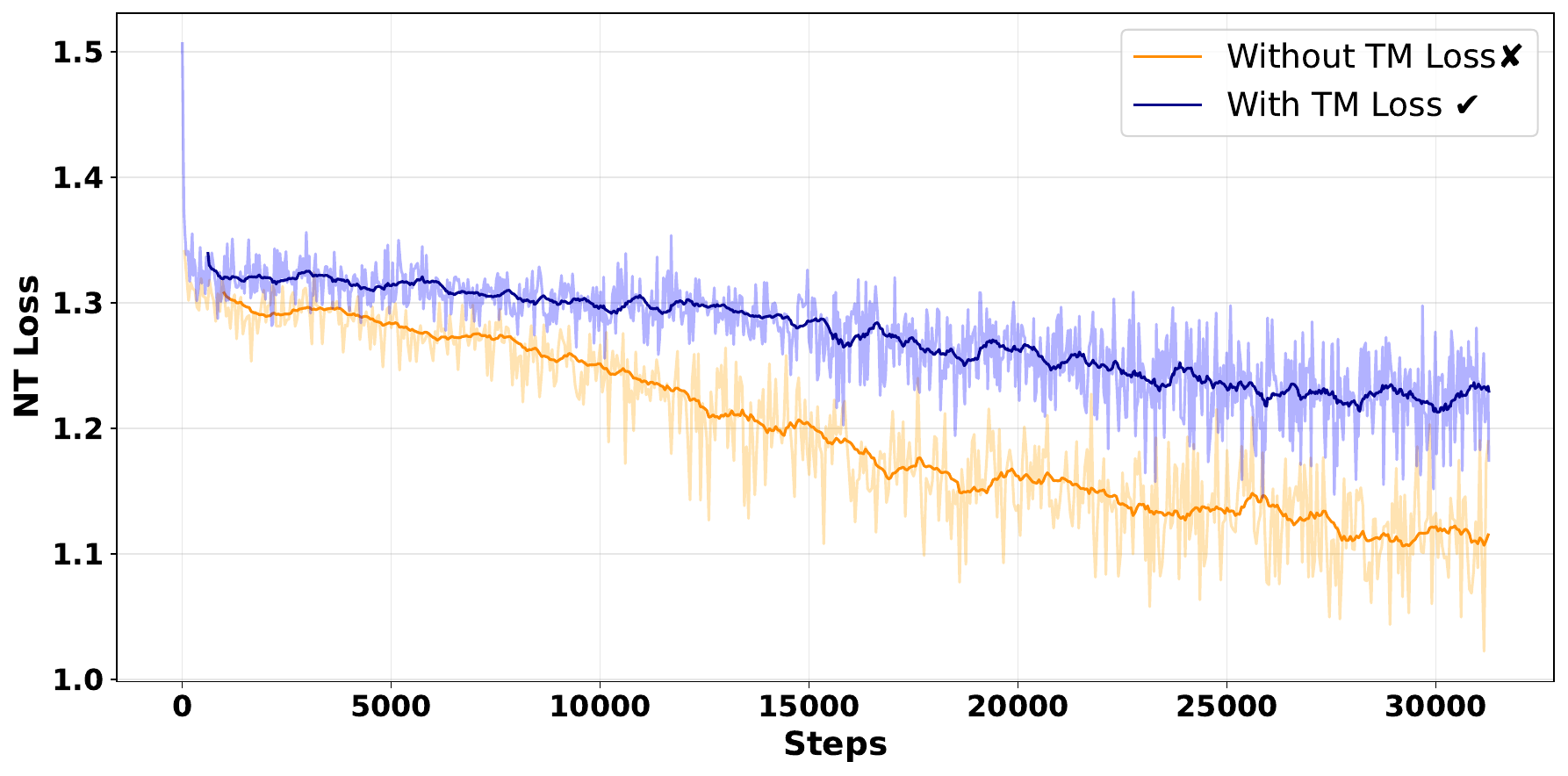} \hfill
  \includegraphics[width=0.48\linewidth]{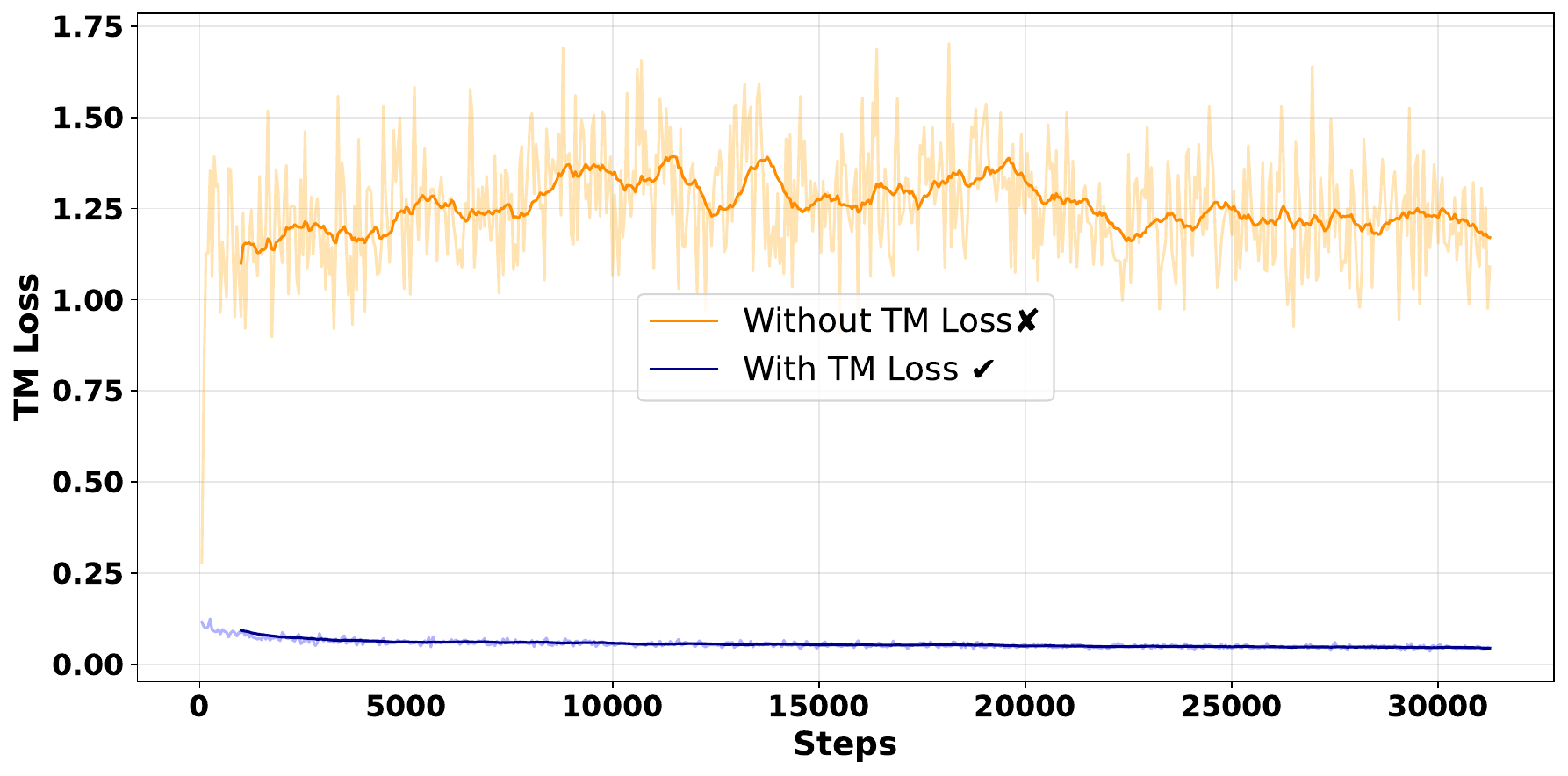}
\caption{Left: Next-token loss; Right: TM loss for HyenaDNA. The convolution-based model struggles to learn the TM effectively, unlike the attention-based model, which better captures Markovian dependencies.}
\label{fig:comp_loss_hyena}
\end{figure}

We trained HyenaDNA on the Scorpio-Gene-Taxa dataset with and without TM loss to investigate its effect on model performance. As shown in Figure \ref{fig:comp_loss_hyena}, the model without TM loss demonstrates better convergence, with the training loss decreasing more effectively compared to the model with TM loss. Notably, the model without TM loss appears to struggle with learning the bigram frequency (transition matrix) inherent to the Markovian model. This observation suggests that convolution-based models, even in an autoregressive manner, may not capture Markovian dependencies as effectively as transformer-based models, which leverage attention mechanisms to model long-range dependencies more efficiently.

\begin{figure}[h]
    \centering
  \includegraphics[width=0.65\columnwidth]{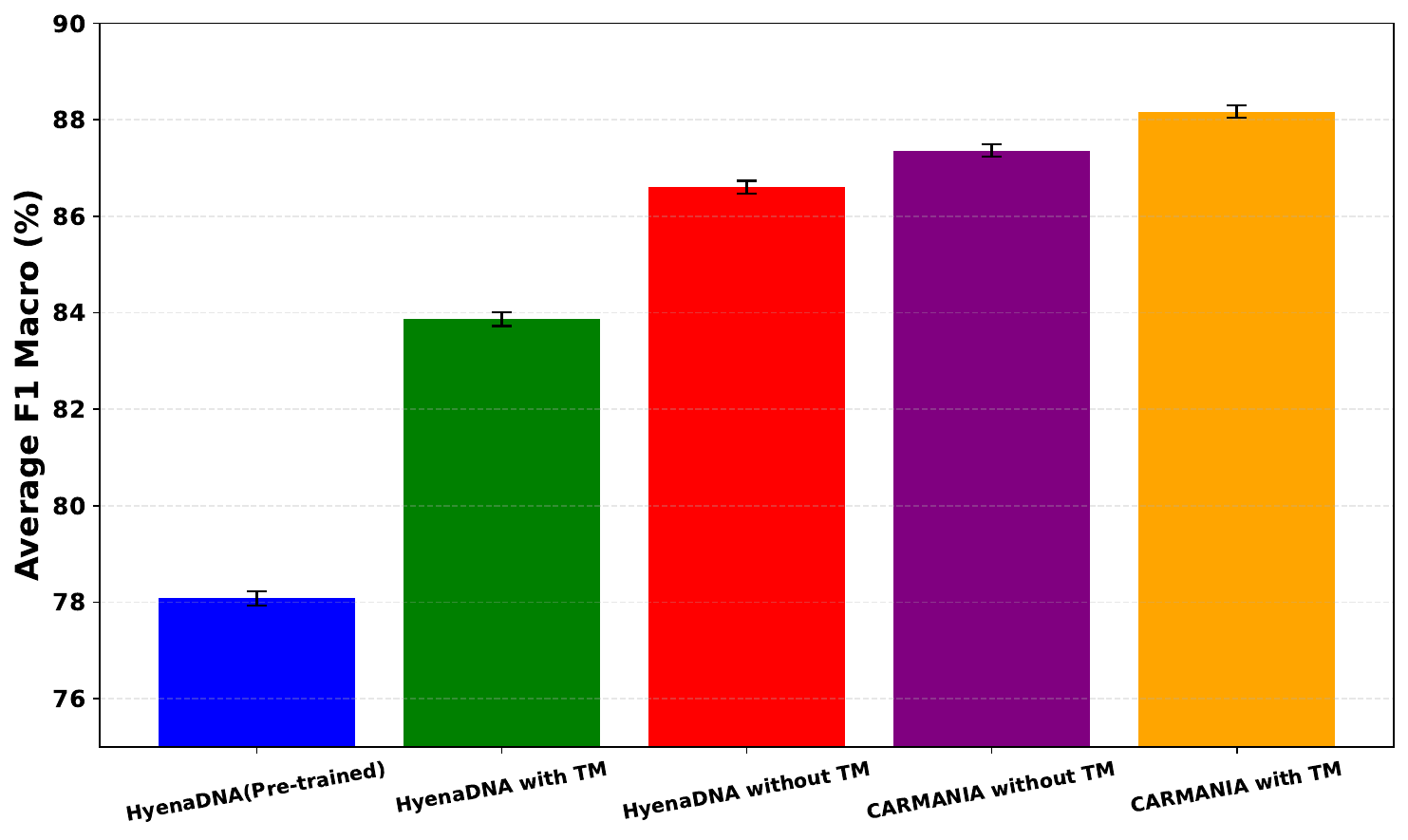}
  \caption{Effect of Transition Matrix Loss on HyenaDNA and \textsc{CARMANIA} : Pre-trained on Scorpio-gene-taxa }
  \label{fig:hyena_tm_effect}
\end{figure}

Additionally, we evaluated the models' performance on the AMR detection tasks using the macro F1 score across all tasks. The results, Figure \ref{fig:hyena_tm_effect},indicate that incorporating TM loss improves performance in CARAMNIA; however, the HyenaDNA model does not exhibit a similar performance gain. This suggests that convolution-based models may inherently struggle to learn embeddings and representations for tasks that rely on Markovian properties, highlighting a potential architectural limitation in such models for sequence modeling tasks.

\subsection{Effect of TM Loss on Mamba-Based Architecture: Caduceus}

{Caduceus}~\cite{schiff2024caduceus} is a bi-directional, reverse-complement equivariant DNA language model built upon the {MambaDNA} architecture, which leverages selective state-space modeling for long-range sequence processing. 
While Caduceus is pretrained using a masked language modeling (MLM) objective distinct from our causal next-token prediction (NT) setup we incorporated the proposed TM loss into the MLM objective while keeping all other configurations unchanged. 
This modification allows us to assess whether TM loss provides additional benefits beyond the CARMANIA backbone.

For this experiment, we used the largest available pretrained variant of Caduceus (approximately 7.7M parameters) and fine-tuned it on the \textbf{Scorpio Gene–Taxa} dataset with and without TM loss. 
Incorporating TM loss led to a notable performance boost, particularly on the gene classification task, yielding a +12\% gain on the test set and +6\% on the held-out set. 
These results suggest that TM loss provides a meaningful signal even in state-space models like Caduceus, although its impact remains more pronounced in transformer-based architectures such as CARMANIA.

\begin{table}[h]
\centering
\caption{
{Effect of TM Loss on Caduceus Fine-Tuned on the Scorpio-Gene--Taxa Dataset.}
Performance (accuracy, \%) across hierarchical taxonomic levels. 
$\Delta$ denotes the relative improvement when adding TM loss.
}
\small
\renewcommand{\arraystretch}{1.05}
\resizebox{0.9\linewidth}{!}{
\begin{tabular}{lccccc|ccccc}
\toprule
 & \multicolumn{5}{c}{\textbf{Test Set}} & \multicolumn{4}{c}{\textbf{Gene-Heldout}}  & \multicolumn{1}{c}{\textbf{Taxa-Heldout}}\\
\cmidrule(lr){2-6} \cmidrule(lr){7-10}  \cmidrule(lr){11-11} 
\textbf{Model} & Phylum & Class & Order & Family & Gene & Phylum & Class & Order & Family & Gene \\
\midrule
\textbf{$\Delta$ (TM $-$ w/o TM)} & +6.2 & +7.0 & +5.8 & +4.0 & +12.4 & +0.3 & +0.1 & $-0.1$ & 0.0 & +6.6 \\
\hline
\textbf{Caduceus w/o TM} & 40.0 & 24.9 & 13.4 & 7.7 & 26.2 & 30.3 & 13.6 & 4.1 & 1.5 & 15.9 \\
\textbf{Caduceus + TM} & 46.2 & 31.9 & 19.2 & 11.7 & 38.6 & 30.6 & 13.7 & 4.0 & 1.5 & 22.5 \\
\midrule
\textbf{CARMANIA} & 86.1 & 76.8 & 59.6 & 41.9 & 90.9 & 46.9 & 31.6 & 15.9 & 9.4 & 72.8 \\
\bottomrule
\end{tabular}
}
\end{table}

\subsection{Sensitivity Analysis: Effect of $\beta$}

The sensitivity analysis evaluates the impact of the hyperparameter $\beta$ on model performance. As shown in Table \ref{tab:parallel_coords}, increasing $\beta$ from 0 to 1.0 improves the macro F1 score (87.36\% to 88.17\%) and BLEU (0.73 to 0.77), suggesting that the TM loss helps capture Markovian dependencies. However, when $\beta$ is increased to 5.0, performance significantly degrades, indicating that excessive TM loss disrupts the next-token objective. The best performance is achieved with $\beta = 1.0$, demonstrating a balanced contribution from both objectives.

\begin{table}[h]
    \centering
    \caption{Comparison of $\beta$ values with F1 Macro, BLEU, and Perplexity scores.}
    \label{tab:parallel_coords}
    \renewcommand{\arraystretch}{1.2}
    \begin{tabular}{c|c|c|c}
        \hline
        \textbf{$\beta$} & \textbf{Average F1 Macro } & \textbf{BLEU} & \textbf{Perplexity} \\
        \hline
        0  &  0.873 & 0.73 & 3.6 \\
        0.5  & 0.874  & 0.76  & 3.6 \\
        1.0  & {0.882}  & 0.77  &3.6 \\
        5.0  & 0.836  & 0.85  & 3.8 \\
        \hline
    \end{tabular}

\end{table}

\subsection{Higher-Order Transition Matrix}

We explored the impact of incorporating TM loss at different Markov orders in \textsc{CARMANIA}, with results shown in Tables~\ref{tab:f1_macro_compare_tm_orders} and~\ref{tab:scorpio_compare_tm_orders}. Surprisingly, while the first-order TM consistently improves downstream classification performance across tasks, the second-order TM performs worse than both the first-order TM and the baseline without TM.

This performance degradation in the second-order setting can be attributed to the increased sparsity of higher-order transition patterns, particularly in biological sequence data. Since second-order transitions require co-occurrence of triplets, many such patterns are infrequent or entirely missing from the training set, leading to unstable or noisy estimations. This sparsity reduces the model's ability to generalize and can hinder the convergence of the transition-based objective\cite{wu2017retrospective}.
On the other hand, first-order transitions strike a balance by enforcing local smoothness and capturing robust pairwise dependencies that are both statistically reliable and biologically meaningful. This results in more stable training and improved generalization across domains.

\begin{table*}[h]
\centering
\caption{Comparison of F1 Macro scores across three AMR classification tasks with different TM orders. First-order TM improves performance across all tasks, while second-order TM leads to degradation.}
\label{tab:f1_macro_compare_tm_orders}
\resizebox{0.6\textwidth}{!}{
\begin{tabular}{lccc}
\hline
\textbf{Model Variant} & \textbf{GeneFamily} & \textbf{Resist-Mech} & \textbf{DrugClass} \\
\hline
Without TM & 0.728	& 0.974 &	0.931 \\
1st-Order TM & \textbf{0.733} & \textbf{0.975} & \textbf{0.942} \\
2nd-Order TM & 0.701 & 0.943 & 0.915 \\
\hline
\end{tabular}}
\end{table*}

\begin{table*}[h]
\centering
\caption{Accuracy comparison on the Scorpio-Gene-Taxa test set across different taxonomic levels. First-order TM leads to consistent improvements, especially at the gene level. Second-order TM degrades performance across all levels.}
\label{tab:scorpio_compare_tm_orders}
\resizebox{0.75\textwidth}{!}{
\begin{tabular}{lccccc}
\hline
\textbf{Model Variant} & \textbf{Phylum} & \textbf{Class} & \textbf{Order} & \textbf{Family} & \textbf{Gene} \\
\hline
Without TM & 0.860 & 0.765 & 0.589 & 0.417 & 0.845 \\
1st-Order TM & \textbf{0.861} & \textbf{0.768} & \textbf{0.596} & \textbf{0.419} & \textbf{0.909} \\
2nd-Order TM & 0.810 & 0.703 & 0.520 & 0.371 & 0.690 \\
\hline
\end{tabular}}
\end{table*}

\subsection{Wide vs. Deep: Architectural Trade-offs for DNA Modeling}

As shown in Table~\ref{tab:deep_vs_wide_model}, our wide architecture reduces the number of layers while significantly increasing the hidden and intermediate sizes. This results in a network that is both more expressive and more efficient.

Most notably, the wide model achieves a substantially higher accuracy on the Scorpio-Gene-Taxa classification task ($71.05\% \pm 18.07$ vs.\ $63.87\% \pm 19.16$), demonstrating its superior ability to capture biologically meaningful patterns from genomic sequences. Furthermore, the wide model also outperforms the deep model on the AMR prediction task—an out-of-domain setting—with a higher macro F1 score ($88.32\% \pm 13.12$ vs.\ $85.22\% \pm 14.66$). These consistent gains across both in-domain and out-of-domain datasets highlight the robustness and generalization ability of the wide configuration.

The observed improvements can be attributed to three key factors:

First, wider layers with large hidden and MLP dimensions better capture motif-level and co-occurrence patterns, which are prevalent in DNA sequences. This capacity is especially critical for tasks such as gene classification and function prediction, where local patterns carry more signal than hierarchical abstractions.

Second, the reduced depth allows for better GPU utilization and faster convergence. With fewer sequential operations, the wide model completes training in nearly half the time while requiring fewer steps to achieve peak performance.

Third, by concentrating representational power within each layer, the wide network is better equipped to model long-range dependencies and subtle biological variations without needing deep hierarchical stacks.
Given these advantages, we adopt the wide model as the backbone for the rest of our study. Its performance and efficiency make it a strong choice for DNA modeling tasks, especially when scaling to large or diverse datasets.

\begin{table}[h]
\centering
\caption{Comparison of model configurations between deep and wide architectures. The wide model uses fewer layers but larger hidden and intermediate sizes.}
\label{tab:deep_vs_wide_model}
\begin{tabular}{lcc}
\toprule
\textbf{Configuration} & \textbf{Deep Model (80M)} & \textbf{Wide Model (83M)} \\
\hline
Number of Layers & 24 & 5 \\
Hidden Size & 512 & 1024 \\
Intermediate (MLP) Size & 1664 & 4608 \\
Attention Heads & 8 & 16 \\
Key/Value Heads & 4 & 4 \\
Attention Window & 128 & 128 \\
Activation Function & SiLU & SiLU \\
\hline
\textbf{Training Steps} & 109,500 & 31,000 \\
\textbf{Epoch Time} & 6.1 hours & 3.2 hours \\
\hline
\hline
\textbf{Task} & \multicolumn{2}{c}{\textbf{Performance Comparison}} \\
\bottomrule
Scorpio-Gene-Taxa Test (Accuracy\%) & $63.87 \pm 19.16$ & \textbf{$71.05 \pm 18.07$} \\
AMR (F1 Macro\%) & $85.22 \pm 14.66$ & \textbf{$88.32 \pm 13.12$} \\
\hline
\end{tabular}
\end{table}

\begin{table}[h]
 \centering
 \caption{Parameter Ranges for Model Training}
 \label{tab:param_ranges}

\begin{tabular}{lc}
   
    \hline
    \textbf{Parameter} & \textbf{Range / Value} \\
    \hline
    Optimizer & AdamW \\
    Batch Size & 1, 19, 35 \\
    Num Epochs & 2 \\
    Learning Rate & 5e-4 \\
    Weight Decay & 0.2 \\
    Adam Epsilon & 1e-6 \\
    Adam Betas & (0.9, 0.999) \\
    Warmup Steps & 400 \\
    LR Scheduler Type & Cosine \\
    Max Grad Norm & 0.85, 2 \\
    \hline
\end{tabular}

\end{table}

\begin{figure*}[t]
    \centering
    \raisebox{1.5ex}{\small \textsc{CARMANIA}} \hspace{0.22\linewidth}
    \raisebox{1.5ex}{\small HyenaDNA} \hspace{0.22\linewidth}
    \raisebox{1.5ex}{\small MetaBERTa(BigBird)} \\
    \includegraphics[width=0.3\linewidth]{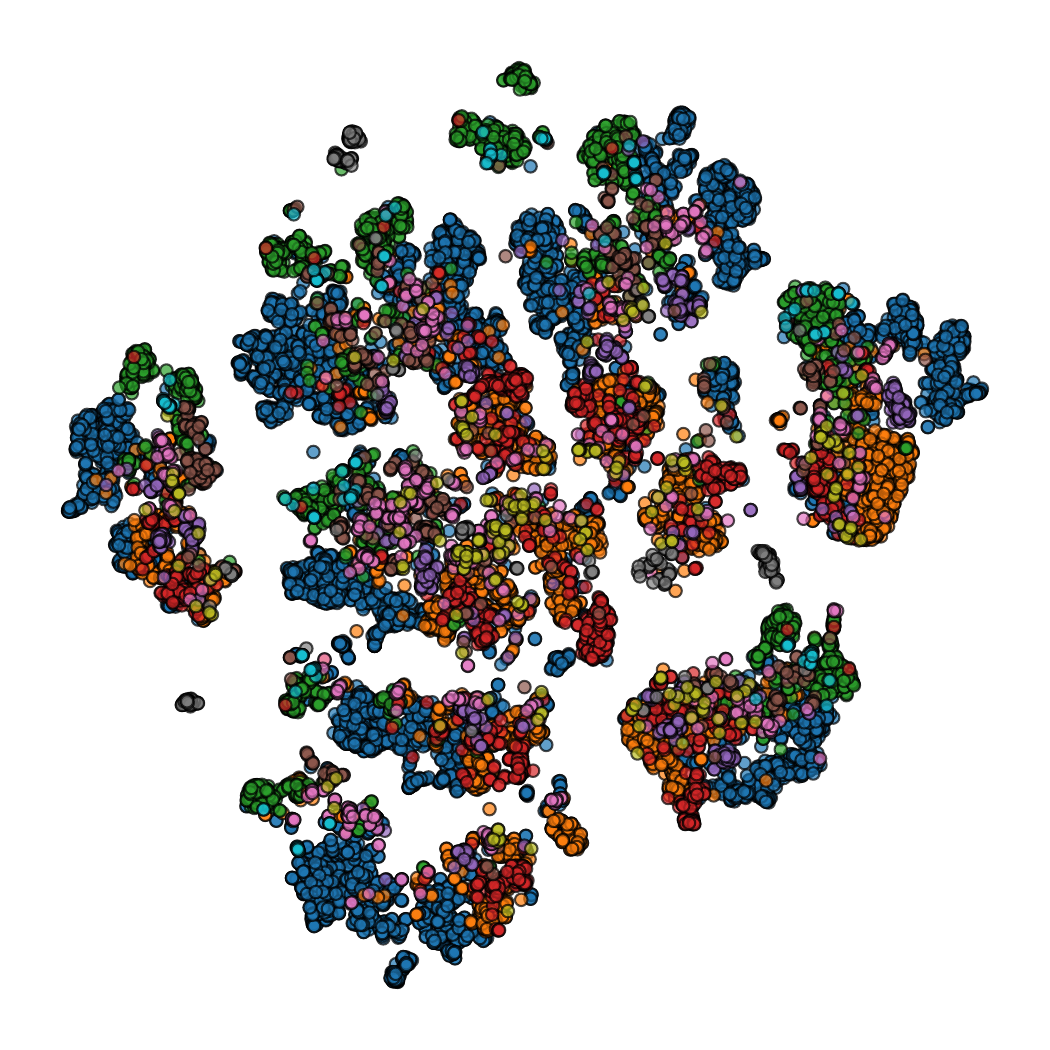} 
    \includegraphics[width=0.3\linewidth]{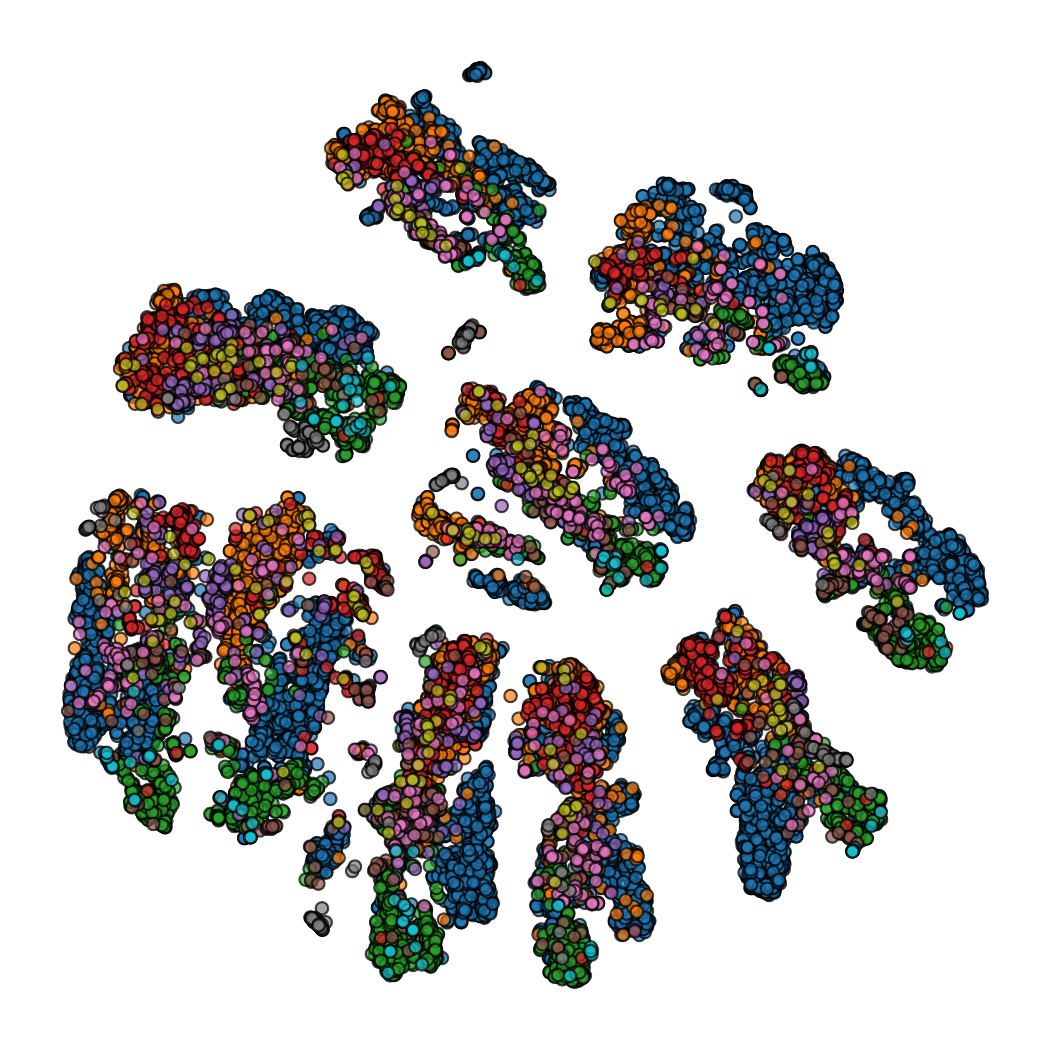}
    \includegraphics[width=0.3\linewidth]{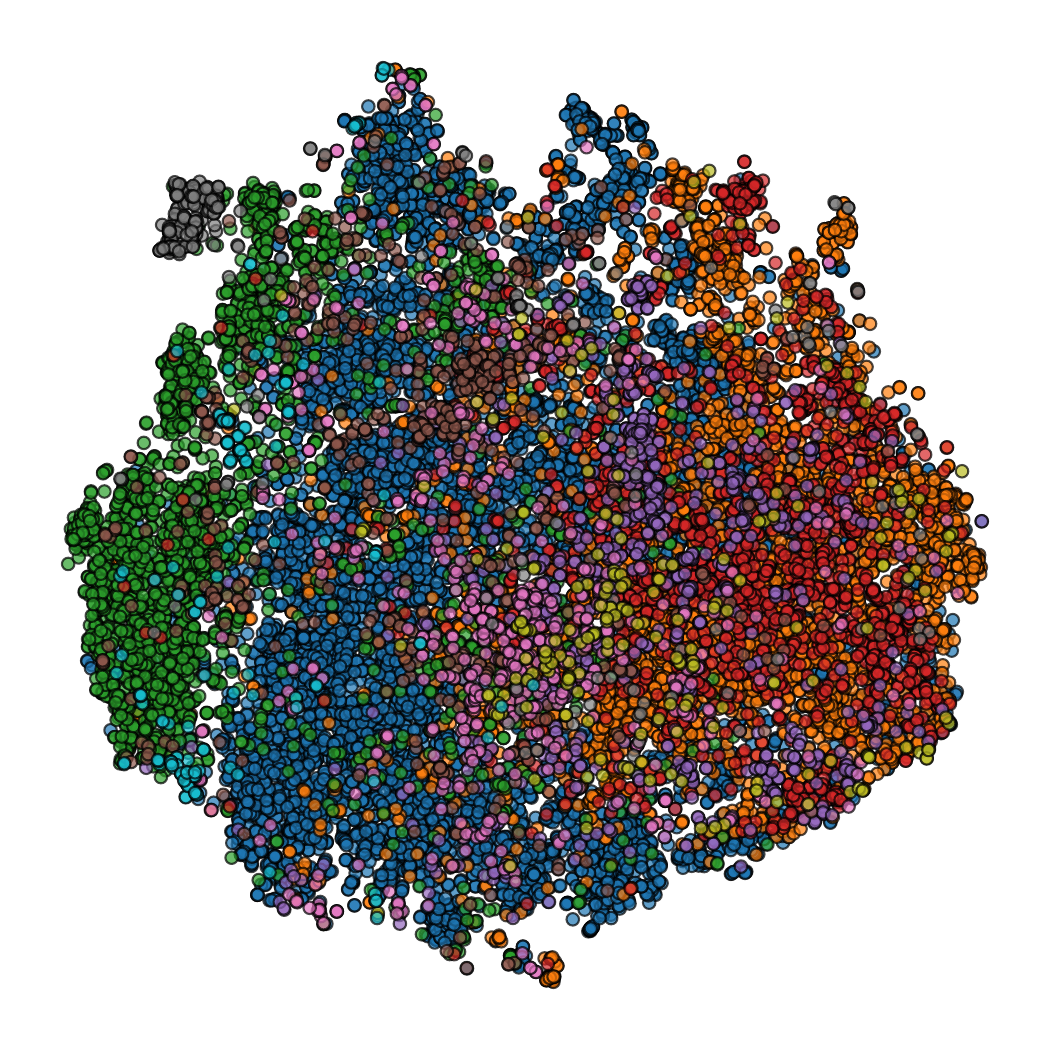}    
    \centerline{\includegraphics[width=1\linewidth]{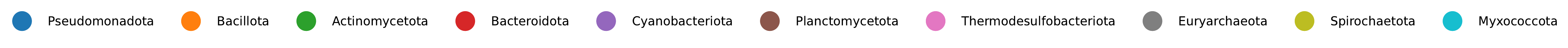}}
    
\caption{t-SNE visualization of the 10 most common phyla in the Scorpio-Gene-Taxa dataset. \textsc{CARMANIA} effectively captures the structure of inter-cluster taxonomy. As shown in Figure \ref{fig:tsne_comparison}, the genes exhibit a clear global structure, with phylogenetic groups positioned near each other when colorized based on phyla. This demonstrates that our model outperforms even MetaBERTa(BigBird) in preserving the taxonomy signal.}
\label{fig:tsne_comparison_phylum}
\end{figure*}

\subsection{Model Size Scalability}

We evaluate the effect of model scaling on performance and compute cost using the Basic Genome dataset. Table~\ref{tab:model_scaling} compares a small 4M-parameter model trained on a 0.5B-token subset with the full 83M-parameter model trained on 10B tokens. Results show substantial gains in both F1 and BLEU scores, demonstrating that increased model capacity and training data size improve representation quality without disproportionately increasing relative FLOPs.

\begin{table}[h]
\centering
\caption{Impact of model size on performance and compute cost.}
\label{tab:model_scaling}
\small
\renewcommand{\arraystretch}{1.0}
\resizebox{0.85\linewidth}{!}{
\begin{tabular}{lccc}
\toprule
\textbf{Model Size / Dataset} & \textbf{F1 Macro(AMR) (↑)} & \textbf{BLEU (Human dataset) (↑)} & \textbf{Relative FLOPs (↓)} \\
\midrule
4M params / 0.5B tokens    & 0.809 ± 0.114 & 0.41 & 0.07 \\
83M params / 10B tokens    & 0.860 ± 0.130 & 0.82 & 1.00 \\
\bottomrule
\end{tabular}
}
\vspace{-6pt}
\end{table}

\subsection{Impact of Sequence Length Scaling}

To evaluate the effect of input sequence length on model performance, we conducted an ablation study using the GRCh38 dataset, comparing models trained on 10~kbp and 160~kbp fragments. As shown in Table~\ref{tab:seq_len_scaling}, increasing sequence length improves both macro F1 and BLEU score, while maintaining the same computational footprint (FLOPs). This suggests that access to longer genomic contexts enables the model to capture more distant dependencies, leading to more robust and informative sequence representations.

\begin{table}[h]
\centering
\caption{Effect of sequence length on performance. Longer input improves both F1 and BLEU without increasing FLOPs.}
\label{tab:seq_len_scaling}
\small
\renewcommand{\arraystretch}{1.0}
\resizebox{0.85\linewidth}{!}{
\begin{tabular}{lccc}
\toprule
\textbf{Training Setup} & \textbf{F1 Macro (↑)} & \textbf{BLEU (↑)} & \textbf{Relative FLOPs (↓)} \\
\midrule
Trained on 10k Sequences   & 0.765 ± 0.120 & 0.82 & 1.00 \\
Trained on 160k Sequences  & 0.800 ± 0.138 & 0.84 & 1.00 \\
\bottomrule
\end{tabular}
}
\end{table}


\subsection{Comparison with Other Convolution-Based Models (ConvNova)}

To contextualize CARMANIA’s performance with recent convolution-based models, we compared it against {ConvNova}~\cite{bo2025revisiting}, a recently proposed architecture for long DNA modeling. 
Following their evaluation protocol, we used five random seeds on the same test benchmarks to ensure consistency. 
As shown in Table~\ref{tab:convnova_comparison}, CARMANIA achieves comparable or superior results on several tasks, demonstrating that its attention-based and Markovian design effectively captures long-range dependencies.

\begin{table}[h]
\centering
\caption{
{Genomics Benchmark Results.}
Top-1 accuracy (↑) is reported for pretrained HyenaDNA, Caduceus-Ph, ConvNova, and CARMANIA, alongside the CNN baseline. 
The best result is in bold, and the second-best is underlined.
}
\label{tab:convnova_comparison}
\small
\renewcommand{\arraystretch}{1.05}
\resizebox{0.9\linewidth}{!}{
\begin{tabular}{lccccc}
\toprule
\textbf{Task} & \textbf{CNN} & \textbf{HyenaDNA} & \textbf{Caduceus-Ph} & \textbf{ConvNova} & \textbf{CARMANIA} \\
\midrule
\textit{Enhancers} \\
\quad Mouse Enhancers & 0.730 ± 0.032 & 0.779 ± 0.013 & 0.754 ± 0.074 & 0.784 ± 0.009 & \textbf{0.795 ± 0.002} \\
\quad Human Enhancers (Cohn) & 0.702 ± 0.021 & 0.718 ± 0.008 & \textbf{0.747 ± 0.004} & \underline{0.743 ± 0.005} & 0.725 ± 0.002 \\
\quad Human Enhancers (Ensembl) & 0.744 ± 0.122 & 0.832 ± 0.006 & {0.893 ± 0.008} & \textbf{0.900 ± 0.004} & \underline{0.895 ± 0.002} \\
\hline
\textit{Species Classification} \\
\quad Coding vs. Intergenic & 0.892 ± 0.008 & 0.904 ± 0.008 & 0.915 ± 0.003 & \textbf{0.943 ± 0.001} & \underline{0.930 ± 0.001} \\
\quad Human vs. Worm & 0.942 ± 0.002 & 0.961 ± 0.002 & \textbf{0.973 ± 0.001} & {0.967 ± 0.002} & \underline{0.970 ± 0.005} \\
\hline
\textit{Regulatory Elements} \\
\quad Human Regulatory & 0.872 ± 0.005 & 0.862 ± 0.004 & 0.872 ± 0.011 & \underline{0.873 ± 0.002} & \textbf{0.894 ± 0.002} \\
\quad Human Non-TATA Promoters & 0.861 ± 0.009 & 0.887 ± 0.005 & {0.946 ± 0.007} & \underline{0.951 ± 0.003} & \textbf{0.965 ± 0.002} \\
\quad Human OCR (Ensembl) & 0.698 ± 0.013 & 0.744 ± 0.019 & \textbf{0.828 ± 0.006} & \underline{0.793 ± 0.004} & 0.778 ± 0.002 \\
\bottomrule
\end{tabular}
}
\vspace{-5pt}
\end{table}

\subsection{Pre-training Dataset}
The pre-training phase utilizes large-scale genomic sequences to establish robust sequence embeddings, ensuring scalability, diversity, and compatibility with existing models trained on domain-specific datasets. We incorporate multiple datasets to support comprehensive learning and facilitate direct performance comparisons.

\noindent\textbf{The Human Reference Genome (GRCh38 - hg38)}: This dataset comprises approximately 3 billion base pairs \citet{grch382013p13}. From this genome, we extracted two non-overlapping sets of sequences to ensure coverage across diverse genomic regions. One set consists of long-range sequences with 160~kbp fragments, while the other contains shorter sequences of 10~kbp fragments.  

\noindent\textbf{The Basic Genome Dataset}: Introduced by \citet{zhu2022phylogeny}, this dataset includes 10 billion base pairs from 4,634 bacterial, archaeal, viral, and eukaryotic genomes. Sequences were extracted as 10~kbp fragments from each genome, ensuring representation across various species and evolutionary lineages.  

\noindent\textbf{The Scorpio Gene-Taxa Dataset}: Developed by \citet{refahi2024scorpio}, this dataset spans 580 million base pairs from 2,046 bacterial and archaeal species, encompassing 497 distinct gene types. To maintain consistency, fragments of 4~kbp were extracted, with shorter sequences padded as needed.  
Table~\ref{tab:pretrain_stats} summarizes the key statistics of pre-training datasets.

\begin{table}[ht]
    \caption{Statistics of the Pre-training datasets.}
    \label{tab:pretrain_stats}
    \resizebox{\linewidth}{!}{%
    \begin{tabular}{lccc}
        \hline
        \textbf{Dataset} & \textbf{Number of Samples} & \textbf{Number of Tokens} & \textbf{Length range (bp)} \\
        \hline
        Human Genome-long \citet{grch382013p13} & 19{,}029 & 3B & 160k \\
        Human Genome-short \citet{grch382013p13} & 303{,}921 & 3B & 10k \\
        Basic Genome \citet{zhu2022phylogeny} & 1{,}010{,}237 & 10B & 10k \\
        Scorpio-Gene-Taxa \citet{refahi2024scorpio} & 547{,}523 & 580M & 114--13{,}227 (4k for training) \\
        \hline
    \end{tabular}%
    }
\end{table}

\subsection{Downstream Datasets}
We used five distinct datasets for fine-tuning and evaluation, covering a range of genomic classification tasks and label spaces. These include regulatory element prediction, gene function classification, taxonomic inference, antimicrobial resistance (AMR) detection, and biosynthetic gene cluster (BGC) prediction. Table~\ref{tab:finetune_stats} provides detailed statistics for each dataset.

\paragraph{AMR:} The Antimicrobial Resistance  classification dataset used in our evaluation is based on the benchmark introduced by Yoo et al.~\cite{yoo2024predicting}. It integrates sequences from two major sources: \textbf{MEGARes}~\cite{bonin2023megares}, a manually curated database of antimicrobial resistance genes annotated by gene family and resistance mechanism; and \textbf{CARD}~\cite{jia2016card}, which provides detailed molecular annotations of AMR genes, mechanisms, and associated drugs.

\paragraph{BGC Dataset.} To evaluate functional classification over extended DNA regions, we used the biosynthetic gene cluster (BGC) classification dataset derived from the MiBiG database~\cite{kautsar2020mibig, liu2022deep}, where each cluster is labeled with a secondary metabolite class. Since BGCs vary in length (average 377k bp), we truncated all sequences to 100k bp for model compatibility.

\paragraph{Nucleotide Transformer Tasks.} We evaluated our model on 18 genomic classification tasks introduced by Dalla-Torre et al.~\cite{dalla2024nucleotide}, covering histone mark prediction, regulatory element annotation, and splice site detection. These tasks span a wide range of sequence-based regulatory functions, offering a comprehensive benchmark for assessing model generalization. 

\paragraph{Genomics Benchmark Tasks.} We evaluated supervised adaptation on the Genomic Benchmarks collection~\cite{grevsova2023genomic}, which includes a range of classification tasks such as regulatory element prediction, enhancer detection, and binary species classification.

\noindent\textbf{The Scorpio Gene-Taxa Dataset:} Developed by \citet{refahi2024scorpio}, this dataset spans 580 million base pairs from 2,046 bacterial and archaeal species, encompassing 497 distinct gene types. To maintain consistency, fragments of 4~kbp were extracted, with shorter sequences padded as needed. The dataset features \textit{Test}, \textit{Gene Out}, and \textit{Taxa Out} splits, carefully designed to assess both memorization and generalization—by holding out specific genes or entire taxonomic groups (e.g., phyla) while preserving hierarchical relevance. This enables robust evaluation of gene-level and taxonomy-level generalization.

\begin{table}[ht]
    \small
    \centering
    \caption{Statistics of the fine-tuning datasets.}
    \label{tab:finetune_stats}
    \resizebox{\linewidth}{!}{%
    \begin{tabular}{lccc}
        \hline
        \textbf{Dataset} & \textbf{Number of Tasks} & \textbf{Classes} & \textbf{Length Range (bp)} \\
        \hline
        Genomics Benchmark~\citet{grevsova2023genomic} & 8 & Binary (one multi-class) & 70--4{,}776 \\
        Scorpio-Gene-Taxa~\citet{refahi2024scorpio} & 10 & Multi-class (497--1{,}929) & 114--13{,}227 \\
        Antimicrobial Resistance Prediction~\citet{yoo2024predicting} & 3 & Multi-class& 211--5{,}274 \\
        Nucleotide Transformer Tasks~\citet{dalla2024nucleotide} & 18 & Binary (one multi-class) & 200--500 \\
        Biosynthetic Gene Cluster Prediction~\citet{kautsar2020mibig} & 1 & 8 & 204--8M (avg 377k; Truncated to 100k) \\
        \hline
    \end{tabular}%
    }
\end{table}

\section{Confirming the Transition Matrix Implementation}

The first-order transition probabilities for the next token given the current token are defined as:

\[
P(w_{i+1} \mid w_i) = [P(A \mid w_i),\; P(T \mid w_i),\; P(C \mid w_i),\; P(G \mid w_i)]
\]

Similarly, for the token after next:

\[
P(w_{i+2} \mid w_{i+1}) = [P(A \mid w_{i+1}),\; P(T \mid w_{i+1}),\; P(C \mid w_{i+1}),\; P(G \mid w_{i+1})]
\]

In our implementation, the bigram transition matrix for position \(i\) is computed as:

\[
\mathbf{T}_i = P(w_{i+1} \mid w_i) \times P(w_{i+2} \mid w_{i+1})^T
\]

Expanding this matrix explicitly:

\[
\mathbf{T}_i =
\begin{bmatrix}
P(A \mid w_i)P(A \mid w_{i+1}) & P(A \mid w_i)P(T \mid w_{i+1}) & P(A \mid w_i)P(C \mid w_{i+1}) & P(A \mid w_i)P(G \mid w_{i+1}) \\
P(T \mid w_i)P(A \mid w_{i+1}) & P(T \mid w_i)P(T \mid w_{i+1}) & P(T \mid w_i)P(C \mid w_{i+1}) & P(T \mid w_i)P(G \mid w_{i+1}) \\
P(C \mid w_i)P(A \mid w_{i+1}) & P(C \mid w_i)P(T \mid w_{i+1}) & P(C \mid w_i)P(C \mid w_{i+1}) & P(C \mid w_i)P(G \mid w_{i+1}) \\
P(G \mid w_i)P(A \mid w_{i+1}) & P(G \mid w_i)P(T \mid w_{i+1}) & P(G \mid w_i)P(C \mid w_{i+1}) & P(G \mid w_i)P(G \mid w_{i+1})
\end{bmatrix}
\]

Summing over the entire sequence of length \(N\), we obtain the predicted (but not yet normalized) bigram matrix:

\[
\mathbf{T}_{Pred} =  \sum_{i=1}^{N} \mathbf{T}_i
\]

To confirm equivalence with the bigram frequency matrix, consider the specific cell \( (T \to A) \):

\[
\mathbf{T}_{Pred}(T \to A) = \sum_{i=1}^{N} P(T \mid w_i) P(A \mid w_{i+1})
\]

If we replace probabilities with actual sequence counts, where \( P(w_i = T) = 1 \) if \( w_i = T \) and 0 otherwise, and normalize each row to represent the transition probabilities, we get:

\[
\mathbf{T}_{Pred}(T \to A) = \frac{\sum_{i=1}^{N} \mathbb{I}(w_i = T) \cdot \mathbb{I}(w_{i+1} = A)}{\sum_{x \in \{A,T,C,G\}} \sum_{i=1}^{N} \mathbb{I}(w_i = T) \cdot \mathbb{I}(w_{i+1} = x)}
\]

This aligns with the actual bigram frequency:

\[
P_{actual}(T \to A) = \frac{\text{count}(T \to A)}{\sum_{x \in \{A,T,C,G\}} \text{count}(T \to x)}
\]

Both matrices are row-normalized, ensuring the sum of each row is 1:

\[
\sum_{x \in \{A,T,C,G\}} \mathbf{T}_{Pred}(T \to x) = \sum_{x \in \{A,T,C,G\}} P_{actual}(T \to x) = 1
\]

Thus, when probabilities are replaced with actual sequence values, the predicted transition matrix \( \mathbf{T}_{Pred} \) is equivalent to the actual bigram frequency matrix \( P_{actual} \), confirming that our method is inherently inspired by bigram frequency calculations while extending them with learnable model probabilities.


\newpage
\section*{NeurIPS Paper Checklist}

\begin{enumerate}

\item {\bf Claims}
    \item[] Question: Do the main claims made in the abstract and introduction accurately reflect the paper's contributions and scope?
    \item[] Answer: \answerYes{} 
    \item[] Justification: We carefully reviewed the abstract and introduction, and they do not contain any claims that are not justified in the paper.

\item {\bf Limitations}
    \item[] Question: Does the paper discuss the limitations of the work performed by the authors?
    \item[] Answer: \answerYes{} 
    \item[] Justification: In the paper, we discuss both the strengths and limitations of our approach in the Experimental Results and Conclusion sections, particularly regarding model generalizability and long-range retention boundaries.

\item {\bf Theory assumptions and proofs}
    \item[] Question: For each theoretical result, does the paper provide the full set of assumptions and a complete (and correct) proof?
    \item[] Answer: \answerYes{} 
    \item[] Justification: We clearly describe the assumptions behind our appraoch and provide a full derivation of the first-order transition tensor and loss in Section 3.1 and 3.2. Due to space constraints, extended derivations and higher-order analysis are included in the supplementary material.

    \item {\bf Experimental result reproducibility}
    \item[] Question: Does the paper fully disclose all the information needed to reproduce the main experimental results of the paper to the extent that it affects the main claims and/or conclusions of the paper (regardless of whether the code and data are provided or not)?
    \item[] Answer: \answerYes{} 
    \item[] Justification:We use publicly available datasets and describe all model architectures, hyperparameters, and training procedures in detail. The code will be released upon acceptance to further support reproducibility.

\item {\bf Open access to data and code}
    \item[] Question: Does the paper provide open access to the data and code, with sufficient instructions to faithfully reproduce the main experimental results, as described in supplemental material?
    \item[] Answer: \answerYes{}
    \item[] Justification: All datasets used are publicly available and cited appropriately. We will release the code and detailed instructions for reproducing the main experimental results in the supplementary material and a public repository upon acceptance.

\item {\bf Experimental setting/details}
    \item[] Question: Does the paper specify all the training and test details (e.g., data splits, hyperparameters, how they were chosen, type of optimizer, etc.) necessary to understand the results?
     \item[] Answer: \answerYes{}
        \item[] Justification: The paper provides detailed descriptions of data splits, optimizer settings, training schedules, and hyperparameters in Section 4 and Appendix A. These include the batch size, learning rate, number of epochs, and evaluation metrics used across all benchmarks.

\item {\bf Experiment statistical significance}
    \item[] Question: Does the paper report error bars suitably and correctly defined or other appropriate information about the statistical significance of the experiments?
    \item[] Answer: \answerYes{}
    \item[] Justification: For datasets with official splits, we follow the provided train/test partitions without additional resampling. For other datasets, we adopt the evaluation protocol from prior work, including 5-fold or 10-fold cross-validation where applicable. Additionally, we report average performance across tasks when appropriate to summarize model generalization, as detailed in Section~4.

\item {\bf Experiments compute resources}
    \item[] Question: For each experiment, does the paper provide sufficient information on the computer resources (type of compute workers, memory, time of execution) needed to reproduce the experiments?
    \item[] Answer: \answerYes{}
    \item[] Justification: We report the GPU type, memory configuration, and training time estimates in Section 4.1 and Appendix~A. These include hardware specifications such as A100 GPUs and per-task training durations to support reproducibility.
    
\item {\bf Code of ethics}
    \item[] Question: Does the research conducted in the paper conform, in every respect, with the NeurIPS Code of Ethics \url{https://neurips.cc/public/EthicsGuidelines}?
    \item[] Answer: \answerYes{} 
    \item[] Justification: The research conducted in the paper conform, in every respect, with the
NeurIPS Code of Ethics.

\item {\bf Broader impacts}
    \item[] Question: Does the paper discuss both potential positive societal impacts and negative societal impacts of the work performed?
    \item[] Answer: \answerYes{}
    \item[] Justification: We briefly discuss the potential applications of our method in genomic discovery and functional annotation, which could support advancements in health, biotechnology, and microbial ecology. We also acknowledge possible risks related to dual-use, privacy (in human genomics), or misuse of predictive models, as noted in the Broader Impact section.
    
\item {\bf Safeguards}
    \item[] Question: Does the paper describe safeguards that have been put in place for responsible release of data or models that have a high risk for misuse (e.g., pretrained language models, image generators, or scraped datasets)?
    \item[] Answer: \answerNA{}
    \item[] Justification: The datasets and models used in this work are not considered high risk for misuse; they are based on publicly available genomic data and do not involve sensitive personal or dual-use information.

\item {\bf Licenses for existing assets}
    \item[] Question: Are the creators or original owners of assets (e.g., code, data, models), used in the paper, properly credited and are the license and terms of use explicitly mentioned and properly respected?
    \item[] Answer: \answerYes{}
    \item[] Justification: All external datasets and models used in this work are properly cited, and we rely exclusively on publicly released assets with appropriate licenses, as noted in the references and Appendix~A.

\item {\bf New assets}
    \item[] Question: Are new assets introduced in the paper well documented and is the documentation provided alongside the assets?
    \item[] Answer: \answerNA{}
    \item[] Justification: This work does not introduce any new datasets, models, or software assets beyond those already publicly available and cited.

\item {\bf Crowdsourcing and research with human subjects}
    \item[] Question: For crowdsourcing experiments and research with human subjects, does the paper include the full text of instructions given to participants and screenshots, if applicable, as well as details about compensation (if any)? 
    \item[] Answer: \answerNA{} 
    \item[] Justification:We do not involve crowdsourcing nor research with human subjects.

\item {\bf Institutional review board (IRB) approvals or equivalent for research with human subjects}
    \item[] Question: Does the paper describe potential risks incurred by study participants, whether such risks were disclosed to the subjects, and whether Institutional Review Board (IRB) approvals (or an equivalent approval/review based on the requirements of your country or institution) were obtained?
    \item[] Answer: \answerNA{} 
    \item[] Justification: The paper does not involve crowdsourcing nor research with human subjects.

\item {\bf Declaration of LLM usage}
    \item[] Question: Does the paper describe the usage of LLMs if it is an important, original, or non-standard component of the core methods in this research? Note that if the LLM is used only for writing, editing, or formatting purposes and does not impact the core methodology, scientific rigorousness, or originality of the research, declaration is not required.
    \item[] Answer: \answerNA{}
    \item[] Justification: Large language models were not used as part of the core methodology or experimental components of this research. Any use of LLMs was limited to minor editing and formatting support and does not impact the scientific contributions.

\end{enumerate}

\end{document}